\documentclass[a4paper,11pt]{article}
\pdfoutput=1 
\usepackage{jcappub} 

\usepackage[T1]{fontenc} 
\usepackage{latexsym}    
\usepackage{graphicx}   
\usepackage{verbatim}   
\usepackage[normalem]{ulem} 
\usepackage[dvipsnames]{xcolor}     
\usepackage{tikz}       
\usepackage{subfigure}  
\usepackage{multirow}
\usepackage{hhline}
\usepackage{url}
\usepackage{hyperref}   
\usepackage{tensor} 
\usepackage{placeins}
\usepackage{amssymb}
\usepackage{enumerate}	
\usepackage{braket}		
\usepackage{physics}
\usepackage{color}
\usepackage{booktabs}

\newcommand{\COMREV}[1]{{}}

\newcommand{\scL}{\mathcal{L}}
\newcommand{\scK}{\mathcal{K}}
\newcommand{\MP}{M_\text{Pl}}
\newcommand{\EPN}{\mathrm{EPN}}
\newcommand{\M}{\mathrm{M}}
\newcommand{\Om}[1]{\Omega_\mathrm{#1}}

\title{\boldmath Cosmological tensions in Proca-Nuevo theory}

\author[a]{Hsu-Wen Chiang,}
\emailAdd{jiangxw[at]sustech.edu.cn}
\emailAdd{b98202036[at]ntu.edu.tw}

\author[b,c,d]{Claudia de Rham,}
\emailAdd{c.de-rham[at]imperial.ac.uk}

\author[a]{Sebastian Garcia-Saenz,}
\emailAdd{sgarciasaenz[at]sustech.edu.cn}

\author[a]{Xue Zhou}
\emailAdd{12332935[at]mail.sustech.edu.cn}

\affiliation[a]{Department of Physics, Southern University of Science and Technology, Shenzhen 518055, China}

\affiliation[b]{Abdus Salam Centre for Theoretical Physics, Imperial College, London, SW7 2AZ, UK}
\affiliation[c]{CERCA, Department of Physics, Case Western Reserve University, 10900 Euclid Ave, Cleveland,
OH 44106, USA}
\affiliation[d]{Perimeter Institute for Theoretical Physics, 31 Caroline St N, Waterloo, ON N2L 2Y5, Canada}

\abstract{We study the cosmological predictions of (extended) Proca-Nuevo theory. This vector-tensor theory enjoys stable homogeneous and isotropic solutions characterized by an effective dark energy fluid, with behavior that ranges from freezing quintessential to thawing phantom-like, serving as a motivated framework to scrutinize the cosmological tensions that affect the standard $\Lambda$CDM model. While the model we consider is sufficiently generic to encompass a large class of field theories, it distinguishes itself from scalar dark energy models (quintessential ones, kinetic ones and non-minimally coupled ones) by the presence of what would be classed as a vector degree of freedom which can be for instance inherited from more generic theories of gravity.
We improve on previous work in several directions: we consider a general one-parameter class of background models; identify a so-called `special' model and analyze observational constraints taking also into account perturbations and making use of wide up-to-date catalogs of datasets including recently released ones. We find that the one-parameter Proca-Nuevo model is preferred over $\Lambda$CDM at $1.5\sigma$ when fitting CMB and BAO data, and at $2.4\sigma$ when further adding low-redshift data. The Hubble tension is alleviated, dropping from $5.8\sigma$ to $2.3\sigma$ (resp.\ $1.5\sigma$) between CMB with (and resp.\ without) BAO data and local measurements. On the other hand, we find that the vector field generically introduces a significant enhancement of the effective Newton constant for natural values of parameters, so that matching the observed matter power spectrum requires a mild amount of tuning to suppress the impact of perturbations. Since, at the background level, Proca-Nuevo is degenerate with other classes of theories, our results are also relevant to a wider range of set-ups including and beyond vector-tensor models.}

\begin{document}

\maketitle
\flushbottom

\section{Introduction}
\label{sec: Introduction}

The resolution of observational tensions in cosmological data has emerged as a major challenge in the current area of precision cosmology. Whether the resolution will lie within systematics,  or truly points towards signs of new physics is yet unclear but explorations along all directions have been flourishing.  Prime among these is the ``Hubble tension'', the $5.3\sigma$ difference in the measurements of the Hubble constant $H_0$ from cosmic microwave background (CMB) data by the \textit{Planck} satellite \cite{Rosenberg:2022sdy} and local observations of Cepheids and SNe type Ia distance ladder by the SH0ES collaboration \cite{Riess:2021jrx}. In addition, a slightly less pronounced but still relevant tension occurs in the $\Om{m0} - S_8$ parameter space,\footnote{$\Om{m0}$ is the matter density parameter, with the subscript ``0'' denoting quantities evaluated at the present time; $S_8 \equiv \sigma_8 \sqrt{\Om{m0} / 0.3}$ measures the matter spectrum's power, with $\sigma_8$ the root-mean-square of matter fluctuations at the scale of $8\,\text{Mpc}/h$; $h \equiv H_0 / (100\,\mathrm{km}\ \mathrm{s}^{-1}\ \mathrm{Mpc}^{-1})$ is the rescaled Hubble constant.} with $\Om{m0} = 0.3158 \pm 0.0064$ and $S_8 = 0.829 \pm 0.011$ from the CMB dataset of \textit{Planck} \cite{Planck:2019nip, Rosenberg:2022sdy, Carron:2022eum}, $\Om{m0} = 0.2962 \pm 0.0095$ and $S_8 = 0.836 \pm 0.035$ from the galaxy clustering dataset of DESI \cite{DESI:2024hhd}, and $\Om{m0} = 0.280^{+0.037}_{-0.046}$ and $S_8 = 0.790^{+0.018}_{-0.014}$ from the galaxy weak lensing dataset of DES Y3 + KiDS-1000 \cite{Kilo-DegreeSurvey:2023gfr}, each in weak tension with each other.

Cosmological tensions are threatening the decades-long reign of the $\Lambda$CDM model. If confirmed, a likely suspect would be the least understood ingredient in the model, i.e.\ the cosmological constant. In fact, even before any anomaly had emerged, cosmologists have long entertained the possibility that a dynamical field might be a more natural explanation for dark energy. It has nevertheless proved challenging to devise well-motivated models that match the successes of $\Lambda$CDM while resolving the tensions, since any modification to the Hubble expansion inevitably alters multiple predictions, often leading to unexpected conflicts; see \cite{Knox:2019rjx,DiValentino:2021izs,Abdalla:2022yfr,Kamionkowski:2022pkx,DiValentino:2025sru,Wolf:2025jlc,Ferreira:2025hwc} for reviews. This aspect has motivated more minimal approaches based on phenomenological models, such as the Chevallier-Polarski-Linder (CPL) parametrization of the dark energy equation of state: $w_{\rm DE}=w_0+(1-a)w_a$. A remarkable finding is that recent data appear to potentially indicate an evolving dark energy, which if confirmed would challenge standard minimally coupled models of quintessence~\cite{Wolf:2025jed} (see also \cite{Lee:2022cyh}). In addition, recent data also favor $w_{\rm DE}<-1$ \cite{DESI:2024mwx} (see however \cite{Wolf:2024eph,Huang:2025som,Efstathiou:2025tie,Gialamas:2025pwv,Colgain:2025nzf}), which for matter minimally coupled to standard Einstein gravity would violate the null energy condition. It is well known that ``phantom'' dark energy ($w_{\rm DE} < -1$) can easily lead to catastrophic consequences \cite{Caldwell:1999ew,Qiu:2007fd}, underscoring the need for a deeper understanding of how such behavior might emerge within consistent theoretical frameworks.

A conceptually minimal description of evolving dark energy is given by a scalar field condensate, as in the well-studied quintessence scenario \cite{Tsujikawa:2013fta}. However, the aforementioned issues with phantom-like behavior indicate that other classes of models may be better motivated, including Galileons with broken shift symmetry \cite{Wolf:2025acj}. Of similar nature, another class of models that has emerged as particularly compelling is the vector-tensor class of gravitational theories, specifically the models that include non-trivial interactions, such as derivative self-couplings or non-minimal gravitational couplings, thus generalizing the Proca theory of a massive vector field. This program, starting with the discovery of the so-called Generalized Proca (GP) theory \cite{Tasinato:2014eka,Heisenberg:2014rta,Hull:2015uwa}, has led to a wealth of interesting results of phenomenological interest in astrophysics (e.g.\ hairy black holes \cite{Heisenberg:2017hwb,Heisenberg:2017xda,Ajith:2020ydz} and destabilization mechanisms \cite{Ramazanoglu:2017xbl,Kase:2020yhw,Garcia-Saenz:2021uyv,Demirboga:2021nrc,Chiang:2025gpa}) and cosmology \cite{DeFelice:2016yws,DeFelice:2016uil,Nakamura:2017dnf,deFelice:2017paw,Geng:2021jso,Savaliya:2025cct}, including in particular the possibility to address the Hubble tension \cite{Heisenberg:2020xak,DeFelice:2020sdq}. Importantly, these models have been shown to be theoretically robust, including a consistent constraint structure \cite{BeltranJimenez:2019wrd}. 

An interesting member of the generalized vector-tensor class of theories is the so-called Proca-Nuevo (PN) \cite{deRham:2020yet}, or its extended version (EPN) \cite{deRham:2021efp}. Several aspects motivate us to study these models in connection with the issue of cosmological tensions. The first is that the theory is characterized by cosmological solutions with an effective dark energy fluid, without the need of a cosmological constant, and this fluid may exhibit either phantom-like or quintessence behaviors. Although this is similar to scalar-tensor models, it should be noted that these features emerge from derivative-type interactions rather than from unusual potentials and may therefore be seen as more natural and robust from a fundamental perspective. Indeed, additional graviton polarizations can in some limit be mimicked by what would be classed as additional scalar and vector degrees of freedom for cosmological purposes and by their very nature, enter with specific classes of derivative interactions and non-minimal coupling to the tensor mode (what can be identified as the standard gravitational part). Such features appear in generic theoretically motivated models of gravity, including those with extra dimensions \cite{Deffayet:2001pu} and finite range gravity \cite{Nicolis:2008in} including massive gravity \cite{deRham:2010ik,deRham:2010kj,deRham:2014zqa}.

An advantage of such models is that the hierarchy between the scale of dark energy (encoded for instance in the graviton mass) and the Planck scale is protected by a non-renormalization theorem, see for instance \cite{Heisenberg:2020jtr,deRham:2021yhr,deRham:2022sdl} for the quantum stability of these models and theoretical consistency against their embedding in a standard and healthy UV completion. This implies that these classes of models are technically natural in the sense that one would not expect quantum corrections to destabilize the system. As a small comment, we may note when comparing with cosmological observations, data favor a relatively large mass of the vector field of about two orders of magnitude larger that the Hubble parameter today. If we were to relate this back to a model of massive gravity, this would correspond to a graviton mass of the order of $\mathcal{O}\left(10^{-31}{\rm eV}\right)$ which is still well within the acceptable range, both observationally \cite{deRham:2016nuf} and theoretically  \cite{deRham:2012ew,deRham:2013qqa}.

In what follows we shall focus our analysis on the EPN class of models. Although EPN is inequivalent to GP, it includes a subset of it and also accommodates a greater number of free functions. This allows for both a more general behavior at the phenomenological level and the possibility that specific choices of coefficients could yield particularly simple models. Indeed, the EPN class includes a so-called `special' model, which boasts interesting features such as a minimal set of free functions and the absence of non-minimal interactions with gravity. In fact, for a natural choice of coefficients, the model leads to a background evolution characterized by the same number of parameters as $\Lambda$CDM. Finally, we note that previous studies of the cosmological predictions of EPN have revealed very promising results, providing a very good fit to late-time data in the case of the no-parameter model just mentioned \cite{Anagnostopoulos:2023pvi,Sudharani:2024qnn}.

These considerations prompt us to revisit the question of confronting EPN with data. As mentioned, previous work was restricted to a particular member of this class of theories and focused exclusively on low-redshift data.\footnote{See also \cite{
Heisenberg:2018mxx,Becker:2020azq,Fell:2024pta} for other studies of cosmological aspects of GP theory and \cite{Bohnenblust:2024mou,Kavya:2025gfw,Sultan:2025dqr} for EPN.} We extend this line of investigation in three directions: we consider a more general subclass of models, leading to a one-parameter modification of the Friedmann equation; restricted to this `special' model, we take into account observational constraints also at the level of perturbations; our analysis makes use of a more complete set of newly released state-of-the-art datasets, including early- and late-universe measurements. Specifically, regarding the last point, we utilize data from \textit{Planck} 2018 low-$\ell$ CMB TTEE \cite{Planck:2019nip}, NPIPE PR4 \textit{Planck} CamSpec high-$\ell$ CMB TTTEEE \cite{Rosenberg:2022sdy}, NPIPE PR4 \textit{Planck} CMB lensing \cite{Carron:2022eyg,Carron:2022eum}, DESI DR1 BAO \cite{DESI:2024mwx,DESI:2024lzq, DESI:2024uvr}, Pantheon+ SNe Type Ia \cite{Brout:2022vxf}, SH0ES local $H_0$ measurements \cite{Riess:2021jrx}, SDSS DR16, SDSS DR7 and 6dF galaxy clustering datasets \cite{eBOSS:2019dcv,Ross:2014qpa,Beutler:2012px}. In particular, the inclusion of the \textit{Planck} dataset enables us to analyze the pros and cons of EPN theory more comprehensively and to establish a more informed comparison against other set-ups.
   
The paper is structured as follows. Sec.~\ref{sec: Setup} presents a review of cosmology in EPN theory, both at the homogeneous and isotropic background level and, in the case of the special EPN model, for linear perturbations. In Sec.~\ref{sec: Data and Analysis setting Methodology} we detail the analysis methodology, the astrophysical and cosmological datasets, and the information criteria used in this work. Sec.~\ref{sec: Results} presents our results of parameter posterior, CMB and matter power spectrum, structure growth, and the tension analysis. Here we also discuss the comparison between the results of EPN and other dark energy candidate models. We summarize and offer final remarks in Sec.~\ref{sec: conclusion}.

\section{Set-up}
\label{sec: Setup}

In this section we present a concise review of EPN theory and its predictions in the context of FLRW cosmology. After general discussions of the theory in Sec.\ \ref{subsec: Extended Proca-Nuevo} and the background equations of motion in Sec.\ \ref{subsec: Background evolution}, we turn our attention to the so-called `special' EPN, a distinguished subclass of models characterized by the absence of non-minimal couplings with gravity. We show in Sec.\ \ref{subsec: EPN minimal} how a particular choice of free functions leads to a one-parameter modification of the Friedmann equation. In Sec.\ \ref{subsec: EPN nonminimal} we also point out that some particular instances of this one-parameter class may obtained more naturally from the general EPN model. Finally, in Sec.\ \ref{subsec: Evolution of the perturbation in the minimally coupled EPN model} we review and generalize the analysis of linear perturbations in the special EPN model, extending the results of \cite{deRham:2021efp} to the one-parameter deformation mentioned above.

\subsection{Extended Proca-Nuevo}
\label{subsec: Extended Proca-Nuevo}

Extended Proca-Nuevo (EPN) is a non-linear vector-tensor theory describing a massive spin-1 field $A_\mu$ coupled to gravity. The full action, including also the Einstein-Hilbert term and standard matter, is given by
\begin{equation}
\label{eq: EPN action}
S = \int {\rm d}^4 x \sqrt{-g} \left( \frac{\MP^2}{2} R + \scL_\EPN + \scL_\mathrm{M} \right)  \,,
\end{equation}
where $\MP$ is the reduced Planck mass and
\begin{equation}
\scL_\EPN = - \frac{1}{4} F^{\mu\nu} F_{\mu\nu} + \Lambda^{4} \left( \hat\scL_0 + \hat\scL_1 + \hat\scL_2 + \hat\scL_3 \right)  \,,
\end{equation}
is the EPN Lagrangian \cite{deRham:2020yet,deRham:2021efp}. The latter is a function of the field strength $F_{\mu\nu}=\nabla_\mu A_\nu-\nabla_\nu A_\mu$ and
\begin{align}
\hat\scL_0 &= \alpha_0 (X)  \,,  \nonumber\\
\hat\scL_1 &= \alpha_1 (X) \scL_1 [\scK] + d_{1}(X) \frac{\scL_1 [\nabla A]}{\Lambda^2}  \,,  \nonumber\\
\hat\scL_2 &= \left( \alpha_2 (X) + d_2 (X) \right) \frac{R}{\Lambda^2} 
+ \alpha_{2,X} (X) \scL_2 [\scK] + d_{2,X} (X) \frac{\scL_2 [\nabla A]}{\Lambda^4}  \,,  \nonumber\\
\hat\scL_3 &= \left( \alpha_3 (X) \scK^{\mu\nu} + d_3 (X) \frac{\nabla^\mu A^\nu}{\Lambda^2}\right) \frac{G_{\mu\nu}}{\Lambda^2} 
- \frac{1}{6} \alpha_{3,X} (X) \scL_3 [\scK] - \frac{1}{6} d_{3,X} (X) \frac{\scL_3 [\nabla A]}{\Lambda^6}  \,,
\label{eq: EPN Lagrangian terms}
\end{align}
which are constructed out of the matrices $\scK^\mu{}_\nu \equiv \left( \sqrt{g^{-1} f[A]} \right)^\mu{}_\nu - \delta^\mu_\nu$ and $(\nabla A)^\mu{}_\nu\equiv g^{\mu\alpha}\nabla_\alpha A_\nu$. Here 
\begin{align}
f_{\mu\nu}[A] = g_{\mu\nu} + 2\frac{ \nabla_{(\mu} A_{\nu)} }{\Lambda^2} + \frac{ \nabla_\mu A_\alpha g^{\alpha \beta}\nabla_\nu A_\beta }{\Lambda^4}  \,,
\end{align}
and
\begin{align}
\scL_n [M] = - \frac{1}{(4-n)!} \epsilon^{\mu_1...\mu_4} \epsilon_{\nu_1...\nu_4} {M^{\nu_1}}_{\mu_1} \cdots {M^{\nu_n}}_{\mu_n} {\delta^{\nu_{n+1}}}_{\mu_{n+1}} \cdots {\delta^{\nu_4}}_{\mu_4}  \,,
\end{align}
for any matrix $M$. We have also introduced in \eqref{eq: EPN Lagrangian terms} the scalar
\begin{align}
X \equiv -\frac{1}{2 \Lambda^2} A_\mu A^\mu  \,,
\end{align}
and $\Lambda$ denoting an energy scale controlling the strength of non-linearity in the theory. The EPN terms \eqref{eq: EPN Lagrangian terms} involve non-minimal gravitational couplings with the curvature scalar $R$ and Einstein tensor $G_{\mu\nu}$, as well as a set of arbitrary functions $\alpha_n(X)$ and $d_n(X)$ (with the notation $\alpha_{n,X}\equiv \frac{\partial}{\partial X}\alpha_n$).

The operators proportional to the functions $d_n$ are part of the GP class \cite{Tasinato:2014eka,Heisenberg:2014rta}, while the ones proportional to the functions $\alpha_n$ are the PN terms \cite{deRham:2020yet}. EPN theory, which combines these two models, provides a consistent, fully non-linear (in fact, non-polynomial) completion of the minimally coupled Proca theory of a massive vector field. From the viewpoint of degree of freedom count, the theory is described by massive spin-1 and massless spin-2 fields up to the Planck scale and thus provides a predictive framework that captures both non-linear and gravitational effects.

A nice consequence of the combination of GP and PN operators in EPN theory is the possibility of tuning the coefficient functions so as to remove the non-minimal couplings with gravity. Indeed, setting $\alpha_2 + d_2=0$, and $\alpha_3 + d_3=0$, we arrive at
\begin{align}
\hat\scL_0^{\rm s} &= \alpha_0 (X)  \,,  \nonumber\\
\hat\scL_1^{\rm s} &= \alpha_1 (X) \scL_1 [\scK] + d_1 (X) \frac{\scL_1 [\nabla A]}{\Lambda^2}  \,,  \nonumber\\
\hat\scL_2^{\rm s} &= \alpha_{2,X} (X) \left( \scL_2 [\scK] - \frac{\scL_2 [\nabla A]}{\Lambda^4} \right)  \,,  \nonumber\\
\hat\scL_3^{\rm s} &= - \frac{1}{6} \alpha_{3,X} (X) \left( \scL_3 [\scK] - \frac{\scL_3[\nabla A]}{\Lambda^6} \right)  \,,
\end{align}
which we will refer to as the `special' EPN theory.

\subsection{Cosmological backgrounds}
\label{subsec: Background evolution}

In this work we focus on a flat Friedmann-Lemaître-Robertson-Walker (FLRW) background
\begin{align}
g_{\mu\nu} {\rm d} x^\mu {\rm d} x^\nu &= - {\rm d} t^2 + a^{2}(t) \delta_{ij} {\rm d} x^i {\rm d} x^j \,,
\end{align}
with scale factor $a(t)$. Homogeneity and isotropy dictate that the vector field must have the form
\begin{align}
A_\mu {\rm d} x^\mu = - \phi(t) {\rm d} t \,,
\end{align}
in terms of a scalar function $\phi$. Variation of the action \eqref{eq: EPN action} yields the following equations of motion:
\begin{align}
\label{eq: Full Friedmann}
H^{2} &= \frac{1}{3 \MP^2} \left( \rho_\M + \rho_\EPN \right)  \,,\\
\label{eq: Full Raychaudhuri}
\dot{H}+H^{2} &= - \frac{1}{6 \MP^2} \left( \rho_\M + \rho_\EPN + 3 P_\M + 3 P_\EPN \right)  \,,\\
0 &=\alpha_{0,X} + 3 \left( \alpha_{1,X} + d_{1,X} \right) \frac{H \phi}{\Lambda^2} + 6 \left[ \left( \alpha_{2,X} + d_{2,X} \right) + \left( \alpha_{2,XX} + d_{2,XX} \right) \frac{\phi^2}{\Lambda^2} \right] \frac{H^2}{\Lambda^2}  \nonumber\\
&\quad - \left[ 3 \left( \alpha_{3,X} + d_{3,X} \right) + \left( \alpha_{3,XX} + d_{3,XX} \right) \frac{\phi^2}{\Lambda^2} \right] \frac{ H^3 \phi}{\Lambda^4}  \,,
\label{eq: Full constraint}
\end{align}
where the first two correspond to the Friedmann and Raychaudhuri equations, respectively, while \eqref{eq: Full constraint} is obtained from the vector field equation. The fact that the latter is a constraint, rather than an evolution equation for $\phi$, is a consequence of the peculiar structure of the EPN Lagrangian. Here $H = \dot a / a$ is the Hubble parameter, with the overscript dot denoting the time derivative, and $\rho_\M$, $P_\M$, $\rho_\EPN$ and $P_\EPN$ are the effective fluid density and pressure from the matter and EPN Lagrangians, respectively, with
\begin{align}
\label{eq: Full EPN density}
\rho_\EPN &= \Lambda^4 \Bigg\{- \alpha_0 + \alpha_{0,X} \frac{\phi^2}{\Lambda^2} + 3 \left( \alpha_{1,X} + d_{1,X} \right) \frac{H \phi^3}{\Lambda^4}  \\
&\quad + 6 \left[ -\left( \alpha_2 + d_2 \right) + 2 \left( \alpha_{2,X} + d_{2,X} \right) \frac{\phi^2}{\Lambda^2} + \left( \alpha_{2,XX} + d_{2,XX} \right) \frac{\phi^4}{\Lambda^4} \right] \frac{H^2}{\Lambda^2}  \nonumber\\
&\quad- \left[ 5 \left( \alpha_{3,X} + d_{3,X} \right) + \left( \alpha_{3,XX} + d_{3,XX} \right) \frac{\phi^2}{\Lambda^2} \right] \frac{H^3 \phi^3}{\Lambda^6} \Bigg\}  \nonumber  \,,\\
\label{eq: Full EPN pressure}
P_\EPN &= \Lambda^{4} \Bigg\{ \alpha_0 - \left( \alpha_{1,X} + d_{1,X} \right) \frac{\phi^2 \dot\phi}{\Lambda^4} + 2 \left( \alpha_2 + d_2 \right) \frac{3 H^2 + 2 \dot{H}}{\Lambda^2}  \\
&\quad- 2 \left( \alpha_{2,X} + d_{2,X} \right) \frac{ \phi \left( 3 H^2 \phi + 2 H \dot\phi + 2 \dot{H} \phi \right)}{\Lambda^4} - 4 \left( \alpha_{2,XX} + d_{2,XX} \right) \frac{H \phi^3 \dot\phi}{\Lambda^6}  \nonumber\\
&\quad+ \left[ \left( \alpha_{3,X} + d_{3,X} \right) \frac{2 H^2 \phi + 3 H \dot\phi + 2 \dot{H} \phi}{\Lambda^3} + \left( \alpha_{3,XX} + d_{3,XX} \right) \frac{H \phi^2 \dot\phi}{\Lambda^5} \right] \frac{H \phi^2}{\Lambda^3} \Bigg\}  \,.\nonumber
\end{align}
For the sake of concreteness, we specify at this stage the functions $\alpha_n(X)$ and $d_n(X)$ as follows:\footnote{Throughout this paper we assume $w_\EPN\neq-1$. As shown in~\cite{Tsujikawa:2025wca} in the setting of GP theory, crossing the ``phantom divide'' with positive dark energy density leads to pathologies, and we expect the same to hold in EPN.}
\begin{align}
\alpha_1   &= - b_1 \tilde X^{p_1} - c_1 \tilde X  \,,\quad
\alpha_2    = \frac{b_2}{3} \tilde X^2 X + \frac{c_2}{2} \tilde X X  \,,\quad
\alpha_{3,X}= b_3 \tilde X^2 + c_3 \tilde X  \,,\quad
\alpha_0    = -c_m \tilde X^{p_0}  \,,  \nonumber\\
d_1        &= - e_1 \tilde X^{p_1} + c_1 \tilde X  \,,\quad
d_2         = \frac{e_2}{3} \tilde X^2 X - \frac{c_2}{2} \tilde X X  \,,\quad
d_{3,X}     = e_3 \tilde X^2 - c_3 \tilde X  \,,
\label{eq: EPN parameterization}
\end{align}
where $\tilde X \equiv \MP^{-2} \Lambda^2 X$ and $b_n$, $c_n$, $e_n$, $p_n$ and $c_m$ are constants, left arbitrary for the time being. We remark that, in the case $p_0=1$, $c_m$ is actually fixed in terms of the mass of the vector particle when considering the theory on Minkowski space ($\phi = 0$): 
\begin{eqnarray}
m^2 \equiv \frac{ \Lambda^4 c_m}{\MP^{2}}\,. 
\end{eqnarray}
The constraint equation and effective density can then be expressed as
\begin{align}
\label{eq: EPN parameterization constraint}
p_0 c_m \left( \frac{\phi^2}{2 \MP^2} \right)^{p_0} &= - 3 p_1 (b_1 + e_1) \frac{H \phi}{\Lambda^2} \left( \frac{\phi^2}{2 \MP^2} \right)^{p_1} + \frac{15 (b_2 + e_2) H^2 \phi ^6}{4 \MP^4 \Lambda^4} - \frac{7 (b_3 + e_3) H^3 \phi^7}{8 \MP^4 \Lambda^6}  \,,\\
\label{eq: EPN parameterization density}
\frac{\rho_\EPN}{\Lambda^4} &= c_m \left( \frac{\phi^2}{2 \MP^2} \right)^{p_0} + \frac{5 (b_2 + e_2) H^2 \phi^6}{4 \MP^4 \Lambda^4} - \frac{(b_3 + e_3) H^3 \phi^7}{2 \MP^4 \Lambda^6}   \,.
\end{align}
This parameterization generalizes the model studied in \cite{deRham:2021efp} by the inclusion of the powers $p_0$ and $p_1$, and it is analogous to the one employed in \cite{DeFelice:2020sdq} in the context of GP theory. Although valid in a phenomenological study, it is clear that not every choice of $p_0$ and $p_1$ will correspond to a consistent effective field theory.

It is worth commenting on the similarities and differences between the cosmology in EPN theory and other cosmological models. The key feature, as already mentioned, is the fact that the scalar $\phi$ satisfies an algebraic equation, as well as the fact that $\phi$ only enters algebraically in the Friedmann equation. In principle, then, one may eliminate $\phi$ algebraically so as to obtained a fully decoupled evolution equation for the scale factor (provided, as usual, that $\rho_\M$ is known in terms of $a$). This is in contrast with scalar-tensor theories including quintessence \cite{Tsujikawa:2013fta} and Horndeski theory \cite{Kreisch:2017uet,Noller:2018wyv,SpurioMancini:2019rxy,Bayarsaikhan:2020jww,Traykova:2021hbr} where the scalar field is dynamical and the equations cannot in general be decoupled (unless a certain global symmetry exists \cite{Capozziello:1996bi}). On the other hand, there exist several other models of dark energy that may be recast into an independent modified Friedmann equation, such as Palatini $f(R)$ \cite{Amarzguioui:2005zq}, $f(T)$ \cite{Benetti:2020hxp,Kumar:2022nvf} and $f(Q)$ theories \cite{BeltranJimenez:2019tme}, holographic dark energy \cite{Wang:2016och}, and some braneworld scenarios \cite{Dvali:2003rk,Binetruy:1999ut,Binetruy:1999hy,Deffayet:2000uy,Deffayet:2001pu}. It is an attractive aspect of EPN theory that this property is achieved in what is arguably a more minimalistic way, in particular without any modification of the gravitational sector, while making explicit the properties of the additional degrees of freedom required to drive the accelerated expansion. 

EPN is however not unique in this respect, since GP theory also enjoys an analogous constraint on the vector field \cite{deFelice:2017paw, DeFelice:2020sdq}. In fact, at the cosmological background level, the constraint of EPN is formally identical to the one of GP, as is clear from the structure of \eqref{eq: Full constraint}. The results of our analysis, when restricted to the background dynamics, will therefore be directly applicable to GP theory (and in fact also to more general theories of gravity with non-minimally coupled non-scalar degrees of freedom, as we will remark later). On the other hand, we emphasize that perturbations break this degeneracy, which as already mentioned is an important motivation for taking them into account in our analysis.

\subsection{Special and very special EPN models}
\label{subsec: EPN minimal}

We recalled in Sec.\ \ref{subsec: Extended Proca-Nuevo} that there exists a choice for the functions $d_2(X)$ and $d_3(X)$ which results in the cancellation of the non-minimal couplings between the vector field and the metric. At the level of the parametrization \eqref{eq: EPN parameterization}, this corresponds to setting $b_2 + e_2 = 0$ and $b_3 + e_3 = 0$. This `special' model is of particular interest as it guarantees unity of the tensor perturbation propagation speed; deviations from the speed of light are severely constrained by the neutron star binary merger observation \cite{LIGOScientific:2017zic} at LVK frequencies. Whether those bound apply to models of dark energy as the ones proposed here depends on the cutoff of the effective field theory \cite{deRham:2018red,Harry:2022zey,Baker:2022eiz}.

Even without non-minimal coupling, the effective energy density of the EPN field possesses non-trivial scaling with respect to $H$,
\begin{align}
\label{eq: General M density}
\rho_\EPN = c_m \Lambda^4 \left( \frac{3 p_1 (b_1 + e_1)}{p_0 c_m} \frac{\sqrt{2}\MP H}{\Lambda^2} \right)^{2M} = \rho_{\EPN,0} \left( \frac{H}{H_0} \right)^{2M}  \,,
\end{align}
where
\begin{equation} \label{eq: M definition}
M \equiv -\frac{p_0}{\left( 1 - 2 p_0 + 2 p_1 \right)} \,,
\end{equation}
is the scaling parameter of the background cosmic evolution. Here we keep $p_0$ as a free parameter for the sake of generality but, as explained previously, one would expect a consistent EFT to have $p_0 = 1$ and $p_1 = 2$, corresponding to the particular model discussed and analyzed in \cite{deRham:2021efp,Anagnostopoulos:2023pvi,Sudharani:2024qnn}. This choice leads to $M = -1/3 < 0$, hinting at the fundamental phantom nature of EPN. To avoid confusion, we refer to this particular case as `very special model', while the special model with $M$ kept as adjustable parameter will be called `special M model'.\footnote{As we discuss in Sec.\ \ref{sec: Data and Analysis setting Methodology}, later we will need to distinguish between models valid only at the background level (neglecting perturbations of the dark energy field) and those which take full account of linear perturbations. The `very special model' refer to the latter.}

We may verify the phantomness property by solving the Friedmann and Raychaudhuri equations, cf.\ \eqref{eq: Full Friedmann} and \eqref{eq: Full Raychaudhuri}, in terms of density parameters $\Omega_i$ and equations of state parameters $w_i$,
\begin{align}
\label{eq: density parameter}
\Omega_i \equiv \frac{\rho_i}{3 \MP^2 H^2}  \,,\quad
w_i \equiv \frac{P_i}{\rho_i}  \,,\quad
1 = \sum_i \Omega_i  \,,\quad
\frac{\dot H}{H^2} = - \frac{3}{2} \sum_i (1+w_i) \Omega_i  \,,
\end{align}
where $i$ runs over individual components such as baryons, cold dark matter, etc.

\begin{figure}
\centering

\subfigure[\label{fig: evolution w}$w_\EPN (z)$]{\includegraphics[width = 0.49 \textwidth]{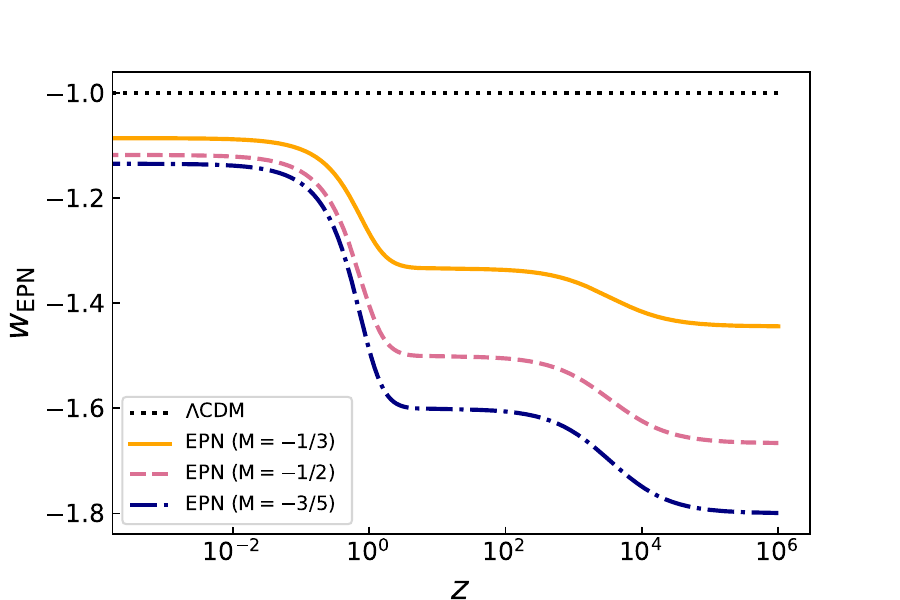}}~
\subfigure[\label{fig: evolution omega}$\Om{EPN} (z)$]{\includegraphics[width = 0.49 \textwidth]{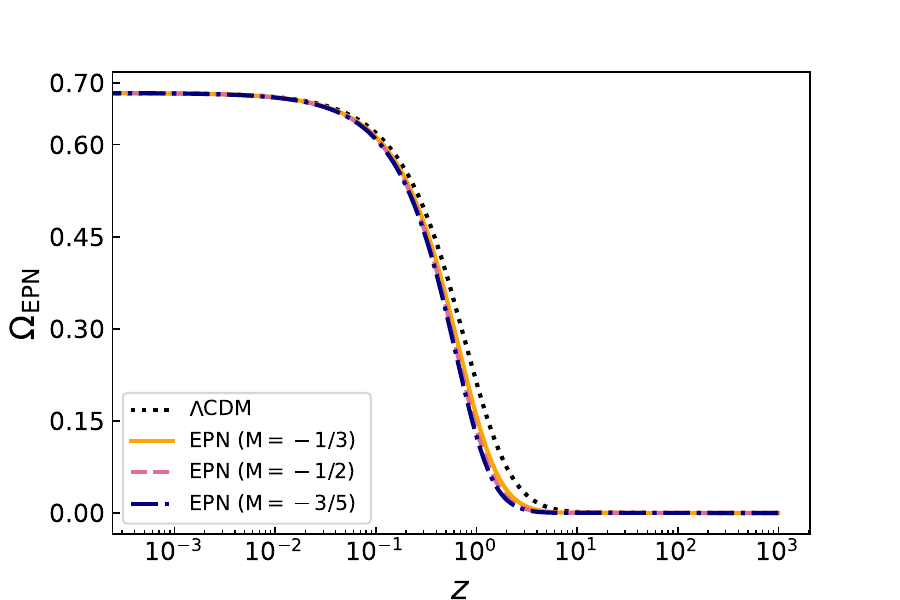}}
\caption{The evolution of the dark energy equation of state parameter $w_\EPN$ (\emph{left panel}) and its energy fraction $\Om{EPN}$ (\emph{right}). The curves display the results in the $\Lambda$CDM model (dotted black), EPN special M model with $M = -1/3$ (solid yellow), and EPN general models with $M = -1/2$ (dashed pink) and $M = -3/5$ (dashed-dotted blue).}
\label{fig: evolution of omega and w}
\end{figure}

For simplicity, let us consider two non-interacting fluids in addition to EPN: radiation `$\mathrm{r}$' representing photons, neutrino kinetic energy, etc., and pressureless matter `$\mathrm{m}$' representing baryons, cold dark matter, etc., with the density scaling $\rho_\mathrm{r} \propto a^{-4}$ and $\rho_\mathrm{m} \propto a^{-3}$. Then we have
\begin{align}
\label{eq: General M Friedmann equation}
\left(\frac{H}{H_{0}}\right)^{2} &= \Om{m0}(1+z)^{3} + \Om{r0} (1+z)^{4} + \Om{EPN0} \left(\frac{H}{H_{0}}\right)^{2M}  \,,
\end{align}
where $z = 1/ a(t) - 1$ is the redshift (with the scale factor at present time $a_0 \equiv 1$). The effective equation of state of the EPN model $w_\EPN$ is then inferred to be
\begin{align}
\label{eq: General M equation of state}
w_\EPN &= - \frac{ 1 - M - M \Om{r} / 3 }{ 1 - M \Om{EPN} } \,.
\end{align}
It follows that $w_\EPN <-1$ if $M<0$, equivalently $p_1 \geq p_0 - 1/2$, evincing phantom behavior across broad ranges of EFT setups without relying on non-minimal gravitational interactions to break the null energy condition. This feature makes the special M model particularly interesting and warrants close inspection. On the other hand, for $0 < M < 1/2$, i.e.\ $p_1 \leq - 1/2$, the system behaves like quintessence, with $-1 < w_\EPN <-1/3$. It is clear however that $w_\EPN$ cannot cross the phantom divide $w = -1$, i.e.\ the so-called quintom scenario, as illustrated in Fig.~\ref{fig: evolution of omega and w}. 
Instead it follows a thawing phantom ($w < - 1$, $\dot w > 0$) or freezing quintessence ($w > -1$, $\dot w < 0$) trajectory with the tracking behavior
\begin{eqnarray}
\label{eq: General M tracking}
w_\EPN =\left\{\begin{array}{lcl}
 - 1 + 4M/3,   & &  \text{in radiation-dominated era}    \,,  \\
 - 1 +  M,  & & \text{in matter-dominated era}       \,,  \\
 - 1,          & &  \text{in dark-energy-dominated era}  \,.  
\end{array}\right.
\end{eqnarray}
This aspect motivates the investigation of more general parameterizations in future work.
The EPN equation of state also constrains the scaling of the condensate field:
\begin{align}
\label{eq: General M phi}
\left( \frac{\phi^2}{\phi_0^2} \right)^{p_0} &= \left(\frac{H}{H_0} \right)^{2M}  \,.
\end{align}
A phantom EPN model implies condensation of the vector in a deceleration phase of expansion and `evaporation' in a phase of acceleration, and vice versa for a quintessence EPN model. The phantom behavior could make one wary of the presence of instabilities. However, as shown in \cite{deRham:2021efp} and as reviewed below in \ref{subsec: Evolution of the perturbation in the minimally coupled EPN model}, there exists a range of parameters in which linear perturbations are fully stable with no indications of strong coupling issues as far as we can identify within this effective theory.

\subsection{General EPN model}
\label{subsec: EPN nonminimal}

In the general EPN model, solving for the scalar component $\phi$ from \eqref{eq: Full constraint} cannot be done in closed form for generic coefficient functions, or even for the generic parametrization of \eqref{eq: EPN parameterization}, cf.\ \eqref{eq: EPN parameterization constraint}. However, particular choices of constants lead to simple closed-form models.

First, by setting $b_1 + e_1=0$, $b_3 + e_3 = 0$ and $b_2 + e_2 \neq 0$, the corresponding model
\begin{align}
\label{eq: EPN density M = -1/2}
\rho_\EPN = \frac{4\left( c_m/3 \right)^{3/2}  \Lambda^4}{\sqrt{5 ( b_2 + e_2 )}} \frac{\Lambda^2}{\sqrt{2}\MP H} = \rho_{\EPN0} \left( \frac{H}{H_0} \right)^{-1}  \,,
\end{align}
permits an algebraic solution to the constraint equation, leading to an EPN energy density equivalent to that of the special M model with $M = -1/2$.

Second, with the choice $b_1 + e_1 =0$, $b_2 + e_2 = 0$ and $b_3 + e_3 \neq 0$ one obtains
\begin{align}
\label{eq: EPN density M = -3/5}
\rho_\EPN &= \frac{22\left(c_m /7\right)^{7/5} \Lambda^4}{\left( b_3 + e_3 \right)^{2/5}} \left( \frac{\Lambda^2}{2\MP H} \right)^{6/5} = \rho_{\EPN0} \left( \frac{H}{H_0} \right )^{-6/5}  \,,
\end{align}
likewise leads to an EPN energy density that mimics the special M model with $M = -3/5$. With this recurring theme, these models prompt us to hypothesize the existence of a family of non-minimally coupled models in EPN with an effective energy density and density fraction parameter that match the same Friedmann equation as in the special M model, Eq.~\eqref{eq: General M Friedmann equation}.

In conclusion, cosmological background models obtained within the special M class discussed in Sec.\ \ref{subsec: EPN minimal} may also be derived as particular instances of the general, non-minimally coupled EPN model. The latter theories may be seen to be more natural from an EFT perspective, although it must be acknowledged that these models still require a tuning of coupling constants. On the other hand, it is worth pointing out that, although the modified Friedmann equations obtained via the two approaches are equivalent, the dynamics of perturbations will certainly differ. Let us also mention again that the one-parameter Friedmann equation thus obtained in EPN coincides also with the one obtained in GP theory, as well as in other very different set-ups such as braneworld models and 
theories of gravity which involve non-minimally coupled higher spin degrees of freedom (i.e.\ of spin $>0$, including vector fields). Any theory of modified gravity using a different derivation formalism or theories akin to $f(T)/f(Q)$ can be recast as gravity with additional non trivial and non-minimally coupled degrees of freedom and can hence exhibit similar features. We note however that while expressing these theories as $f(T)/f(Q)$ may be convenient to derive the classical background equations of motion, being able to identify the relevant degrees of freedom is essential not only to compare with observations but also to ascertain the stability and consistency of these models.
Our analysis, which utilizes a broad and up-to-date catalog of datasets, is therefore applicable to all these models when restricted to the background evolution but goes beyond that.

\subsection{Linear perturbations in the special EPN model}
\label{subsec: Evolution of the perturbation in the minimally coupled EPN model}

Given the aforementioned degeneracy of models at the background level, it is important to understand how perturbations affect the predictions of the theory. In the general EPN theory, the task is challenging due to the complexity of the equations, so for simplicity we consider here the special M model restricted to the case $p_0=1$. Our main aim here is to establish the stability criteria for perturbations, extending the analysis of \cite{deRham:2021efp} that focused on the case $p_1=2$.

Metric perturbations in the flat gauge are decomposed into two scalar modes $\alpha$ and $\chi$, one transverse vector mode $V_i$ and the transverse and traceless tensor mode $h_{ij}$,
\begin{align}
g_{\mu \nu} \mathrm{d} x^\mu \mathrm{d} x^\nu &= - \left( 1 + 2 \frac{\alpha}{\MP} \right) \mathrm{d} t^2 + \frac{2}{\MP} \left( \frac{ \partial_i \chi}{\MP} + a V_i \right) \mathrm{d} t \mathrm{d} x^i+a^2 (t) \left( \delta_{ij} + \frac{h_{ij}}{\MP} \right) \mathrm{d} x^i \mathrm{d} x^j \,.
\end{align}
The perturbed vector field is parametrized by two scalar perturbations $\delta \phi$ and $\chi_V$ along with a transverse vector mode $Z_i$,
\begin{align}
A^0 = \phi(t) + \delta \phi  \,,\quad
A^i = \frac{1}{a^2} \delta^{ij} \left( a Z_j - \frac{a}{\MP} \phi V_j + \frac{\partial_j \chi_V}{\Lambda} \right) \,.
\end{align}
The spatial components of the four-velocity of a perfect fluid matter species can be split as
\begin{align}
\label{4velocity}
u_i = 
-\frac{\partial_i v}{\MP^2} + \frac{U_i}{\MP}  \,,
\end{align}
where $v$ is a scalar and $U_i$ is a transverse vector. Conservation of the energy-momentum tensor gives the continuity and Euler equations describing the time evolution of the $i$-th fluid density perturbation $\delta \rho_i$ and the matter momentum density $(\rho_i + P_i) v_i$,
\begin{align}
\MP^2 \dot{\delta \rho_i} + 3 \MP^2 H \left( 1+c_i^2\right)\delta \rho_i + \frac{k^2}{a^2} \left( \rho_i + P_i \right) v_i + \frac{k^2}{a^2} \left( \rho_i + P_i \right) \chi &= 0  \,,\\
\left( \rho_i + P_i \right) \dot v_i - 3 H c_i^2 \left( \rho_i + P_i \right) v_i - \MP^2 c_i^2 \delta \rho_i  - \MP\left( \rho_i + P_i \right) \alpha &= 0  \,,
\end{align}
with $\delta P_i$ representing the pressure perturbation and the fluid speed of sound is defined via $\delta P_i \equiv c_i^2 \delta \rho_i$. Combined with the EPN equations, it becomes evident that the EPN field influences the evolution of matter, affecting for instance matter aggregation and structure formation. The resulting equations of motions for the scalar sector,\footnote{We refer the reader to \cite{deRham:2021efp} for details on the calculations that follow. We also ignore from now on vector and tensor perturbations. In the special model that we focus on in our perturbation analysis, tensor modes evolve exactly as in GR and therefore do not give any additional constraints. Moreover, matter (assuming a perfect fluid) and metric vector modes are decoupled from those of the EPN field, so the vector sector also does not affect the confrontation of the model with data.}
\begin{align}
&\left( 3 H \hat\omega_1 - 2 \hat\omega_4 \right) \frac{\delta\phi}{\phi} - 2 \hat\omega_4 \frac{\alpha}{\MP} + \sum_i \delta \rho_i + \frac{k^2}{a^2 \Lambda^2} \left[ \hat{\mathcal{Y}} + \hat\omega_1 \frac{\Lambda^2}{\MP^2} \chi - \hat\omega_6 \Lambda \psi \right] = 0  \nonumber  \,,\\
&\MP^2 \hat\omega_2 \frac{\delta \phi}{\phi} + \MP \hat\omega_1 \alpha + \sum_i (\rho_i + P_i) v_i = 0 
 \nonumber  \,,\\
&(3 H \hat\omega_1 - 2 \hat\omega_4 ) \frac{\alpha}{\MP} - 2 \hat\omega_5 \frac{\delta \phi}{\phi} + \frac{k^2}{a^2 \Lambda^2} \left[ \frac{1}{2} \hat{\mathcal{Y}} + \hat\omega_2 \frac{\Lambda^2}{\MP^2} \chi - \frac{\Lambda}{2} ( \hat\omega_2 + \hat\omega_6 \phi ) \frac{\psi}{\phi} \right] = 0  \nonumber  \,,\\
&\frac{\dot{\hat{\mathcal{Y}}}}{H} + \left( 1 - \frac{\dot\phi}{H \phi} \right) \hat{\mathcal{Y}} + \frac{\Lambda^2}{H} \left[ \hat\omega_2 \frac{\delta \phi}{\phi} + 2 \hat\omega_7 \frac{\phi \psi}{\Lambda} + \hat\omega_6 \left( 2 \frac{\alpha \phi}{\MP} + \delta\phi \right) \right] = 0  \label{eq: EOM scalar} \,,
\end{align}
with the definitions
\begin{align}
\psi \equiv \chi_V + \frac{\Lambda}{\MP^2} \phi \, \chi  \,,\quad
\hat{\mathcal{Y}} \equiv \Lambda^2 \hat\omega_3 \left( \frac{\delta\phi}{\phi} + 2 \frac{\alpha}{\MP} + \frac{\dot\psi}{\Lambda\phi} \right)  \,,  \nonumber\\
\hat\omega_1 = \hat\omega_2 - 2 \MP^2 H  \nonumber  \,,\quad
\hat\omega_2 = -2 \MP^2 H \Om{EPN} = \phi \, \hat\omega_6  \nonumber  \,,\quad
\hat\omega_3 = -2 \phi^2 \hat{q}_V  \nonumber  \,,\\
\hat\omega_4 = - 3 \MP^2 H^2 \left[ 1 - \left( 2 p_1 - 3 \right) \Om{EPN} \right]  \nonumber  \,,\quad
\hat\omega_5 = \frac{3}{2} \left( 1-2 p_1  \right) H \hat\omega_2  \nonumber  \,,\quad
\hat\omega_7 = \dot\phi \, \phi^{-3} \hat\omega_2  \nonumber  \,,\\
\hat q_V = 1 - \left( 2 + \frac{\dot\phi + H \phi}{\Lambda^2} \right)^{-1} \left[ \alpha_1 - 2 \left( 1 - 2 \frac{H \phi}{\Lambda^2} \right) \alpha_{2,X} + \frac{H \phi}{\Lambda^2} \left( 2 - \frac{H \phi}{\Lambda^2} \right) \alpha_{3,X} \right]  \,,
\label{eq:coefsomegaspec}
\end{align}
determine the two metric perturbations $\alpha$ and $\chi$, the constraint on $\delta\phi$, and the evolution of the matter perturbation and $\psi$, which are dynamical degrees of freedom.

The absence of ghost, gradient and tachyon instabilities for the EPN and matter perturbations near the de Sitter fixed point ($\Omega_{\rm EPN}\to1$) translate into the conditions of positivity of the following quantities \cite{deRham:2021efp}:
\begin{align}
\hat q_V &=
1- \frac{ b_1  \tilde c_m^{-p_1/2} + ( 10 b_2 + 8 b_3 ) \tilde c_m^{-1} + (c_1 + 10 c_2 + 8 c_3) \tilde c_m^{-1/2}}{\left( 1 + p_1^{-1} / 2
\right) \left(  1 - \Om{EPN} \right) }  \,,\\
\hat c_V^2 \hat q_V &= 
1 - \frac{1}{2} b_1 \tilde c_m^{- p_1 / 2} - (3 b_2 + b_3) \tilde c_m^{-1}  - \left(\frac{c_1}{2} + 3 c_2 + c_3 \right) \tilde c_m^{-1/2}  \,,\\
\hat m_V^2 H^{-2} &= 5 \,,\\
\hat q_{S,\psi} &= \frac{3 \Lambda^2}{\MP^2}\frac{(\Om{EPN} - M^{-1}) \tilde c_m}{(1-\Om{EPN})^2}  \label{eq: q_S}  \,,\\
\hat c_{S,\psi}^2 &= \frac{\tilde c_m^{1/2}}{6 p_1 \hat q_V}  \,,\\
\hat m_{S,\psi}^2 H^{-2} &= \frac{45}{4} - \frac{6 \left( 1 + p_1^{-1} \right)^2 \tilde c_m^{1/2}}{\hat q_V ( 1 - \Om{EPN} )} - \frac{\tilde c_m}{\hat q_V^2 \left( 1 - \Om{EPN} \right)^2}  \,, \label{eq:mspsi} \\
\hat m_{S,\M}^2 H^{-2} &= \frac{3 \tilde c_m^{1/2}}{2 \hat q_V ( 1 - \Om{EPN} )}  \,,
\end{align}
where $\tilde c_m \equiv c_m/6$. Here the $\hat{q}$ parameters represent the coefficients of the kinetic terms of the perturbations, the $\hat{c}_S$ represent the respective speed of sounds, and the $\hat{m}_S$ represent the effective masses; in particular, $\hat m_{S,\M}$ is the effective mass of the matter perturbation in the scalar sector, assumed here to be given by a single perfect fluid (a good approximation during the matter- and dark energy-domination eras, including the transition between the two). We also emphasize that these results are valid up to order $(1-\Omega_{\rm EPN})^0$ in an expansion around the fixed point.

Stability under ghost-, gradient- and tachyon-type solutions demands the positivity of all these quantities. Actually, $\hat m_{S,\psi}^2$ has a negative-definite contribution at order $(1-\Omega_\EPN)^{-1}$, which we omitted as it can be canceled by setting
\begin{equation}
b_1 + e_1=p_1^{-1} \tilde c_m^{(3 - M^{-1})/4} \,,
\end{equation}
which has already been imposed in Eq.\ \eqref{eq:mspsi}.

A simple and convenient choice that meets the stability criteria under ghosts and gradients is
\begin{equation}
b_1=c_1=b_2=c_2=b_3=c_3=-1\,,
\end{equation}
which we will adopt in the numerical analysis that follows. Taking into account Eq.~\eqref{eq: General M density}, this leaves us with two free parameters: the coefficient $M$ of the Friedmann equation and the mass $m$ of the Proca field (equivalently, the constants $p_1$ and $c_m$).

We remark that the above choice does not entail a severe loss of generality: the perturbation parameters enter in the equations only through $\hat\omega_3 = -2\phi^2 \hat q_V$, and the form of $\hat q_V$ ensures that one can always write its numerator as a generic function of $\phi$. Assuming large $c_m$ (an assumption to be justified later based on the results of the data fitting), we have that $\hat q_V-1$ is well approximated by a linear combination of the coefficients $b_i$ and $c_i$ times a single power of $c_m$. As a result, under the assumption of stability, the $b_i$ and $c_i$ parameters are degenerate with $c_m$.

This set-up differs significantly from the GP model studied in \cite{DeFelice:2020sdq}, in which $\hat m_{S,\psi}^2 H^{-2} \sim -2 |e_1| \tilde c_m^{-1/2} \left( 1 - \Om{EPN} \right)^{-2}$ is negative definite. We also remark that the quintessence-like model ($M > 0$) develops a ghost instability as $\hat q_{S,\psi}$ diverges toward $-\infty$ according to Eq.~\eqref{eq: q_S}. Coincidentally, it is also observationally disfavored at the background level, as discussed in Sec.\ \ref{sec: Results}.

As $c_m$ (and $b_i$, $c_i$) only appears in $\hat\omega_3$ in the master equations of the scalar perturbation, Eqs.~\eqref{eq: EOM scalar}, we see that we can achieve the effective decoupling of the EPN scalar mode from ordinary matter in the limit $\hat\omega_3 \propto \hat q_V \to 0$. If $b_i,c_i=\mathcal{O}(1)$, this in turn implies $c_m\propto m^2 \to \infty$ (remembering that we set $p_0=1$), i.e.\ a very heavy field, making it reasonable that perturbations should be suppressed in this limit. We will verify this explicitly in our analysis. Related to this, we also point out that $c_m$ is introduced in the first place precisely because one expects it to be $\mathcal{O}(1)$ in a consistent EFT, where one has $\Lambda\sim (m\MP)^{1/2}$. We will further comment on this point in Sec.\ \ref{sec: conclusion}.

\section{Data and analysis methodology}
\label{sec: Data and Analysis setting Methodology}

In this section, we introduce the analysis methodology and the astrophysical and cosmological datasets to investigate the EPN models described in Section~\ref{sec: Setup}. We focus on two particular realizations of the EPN model. First, we consider the special M model with $M$ ranging from $-1/2$ (phantom) to $1/2$ (quintessence), while ignoring the effect of perturbations of the dark energy fluid entirely. Formally, this amounts to setting $\hat\omega_2 = \hat\omega_3 = \hat\omega_6 = 0$, $\hat\omega_4 = -3 \MP^2 H^2$ and $\hat\omega_1 = - 2 \MP^2 H$ in Eqs.~\eqref{eq: EOM scalar}, resulting in the decoupling of the EPN perturbations from the matter and metric ones, thus reducing the perturbation equations back to the $\Lambda$CDM set-up, with the background evolution substituted by Eq.~\eqref{eq: General M Friedmann equation}. One expects this decoupled theory to be a suitable approximation for describing most high-redshift observations, since $|\hat\omega_2| \ll \MP^2 H$ (cf.\ \eqref{eq:coefsomegaspec}) during the matter-dominated era when the large-scale structure formed. By the same token, we expect this description to break down at low redshifts, and indeed we will see that perturbations do have an important effect on the structure growth rate. We also do not expect this approximation to be valid for the quintessence-like models ($0 < M < 1/2$), since we have seen that perturbations are unstable near the de Sitter fixed point in this set-up. Nevertheless, we still include this parameter range in our analysis, for the sake of completeness on the one hand, but also because of the potential applicability to other theories.

In the second realization we extend the analysis to take full account of perturbations, focusing on their effect on observables such as the matter power spectrum and the growth history. Let us clarify the terminology we employ for the different model realizations: 
\begin{itemize}
\item \textbf{EPN special M model}: This refers to the realization mentioned above in which we ignore the EPN perturbation, either fixing $M$ or allowing it to vary in the prior range $[-0.5,0.5]$;
\item \textbf{EPN very special model}: This is the model discussed in Sec.\ \ref{subsec: EPN minimal}, i.e.\ the EFT-motivated member of the special EPN class with $M=-1/3$, taking into account all scalar perturbations;
\item \textbf{EPN full special model}: This is the special EPN model, taking all scalar perturbations into account, and without fixing $M$. We let the latter vary in the prior range $[-1/2,-10^{-8}]$ (recall that EPN perturbations are stable only for $M<0$).\footnote{The case $M=0$ yields ill-defined perturbations equations, which is why we exclude this value from our prior. The reason can be traced back to the definition of $M$ in terms $p_0$ and $p_1$, Eq.\ \eqref{eq: M definition}: having fixed $p_0=1$, $M$ must be strictly non-zero.}
\end{itemize}
Furthermore, for the latter two cases we choose the prior range $[-1,6]$ for $\log_{10} c_m$. As discussed in the previous section, sampling over parametrically large values of $c_m$ will allow us to explore the decoupling regime of EPN perturbations.

We evaluate the models with the Boltzmann code \texttt{CAMB} \cite{Lewis:1999bs,Howlett:2012mh}. For the $\Lambda$CDM model, the standard six-parameter basis is used: the physical densities of baryons $\Om{b0} h^2$ and cold dark matter $\Om{c0} h^2$, the approximated acoustic angular scale $\theta_{\rm MC}$, the optical depth $\tau_{\rm reio}$, the amplitude of primordial scalar perturbations $\ln ( 10^{10} A_{\rm s} )$, and the scalar spectral index $n_{\rm s}$. The full special EPN model is implemented in \texttt{CAMB} with the additional parameters $M$ and $c_m$.\footnote{The scalar perturbation equations, Eqs.~\eqref{eq: EOM scalar}, are transported to \texttt{CAMB} variables via the following mapping:
\begin{align}
\frac{\alpha}{\MP} = \frac{ H \partial_t\text{etak} - \text{etak} \dot H}{k H^2}  \,&,\quad
\frac{\sum_i \delta\rho_i}{\MP^2} = \frac{\text{dgrho}_\text{noDE}}{a^2} - 3 \frac{\rho_\M}{\MP^2}  ( 1 + w_\M ) \frac{\text{etak}}{k}  \,,\nonumber\\
\frac{\chi}{\MP^2} = a \frac{z}{k} - \frac{\text{etak}}{k H} + \frac{3 a^2 \partial_t \text{etak}}{k^3}  \,&,\quad
\frac{\sum_i (\rho_i + P_i) v_i}{\MP^4} = \frac{\text{dgq}_\text{noDE}}{a k} + \frac{\rho_\M ( 1 + w_\M )}{\MP^2 H} \frac{\text{etak}}{k}  \,.\nonumber
\end{align}
}
In view of the modifications to the matter perturbation evolution, we have preliminarily considered both the analytic \texttt{halofit}~\cite{Smith:2002dz,Seljak:2000gq} and iterative \texttt{HMcode}~\cite{Mead:2020vgs} halo models for the nonlinear matter power spectrum. Having checked that both produce identical results given randomly drawn parameters (and provided \texttt{HMcode} converges, as indeed it does except for rare edge cases), we have picked \texttt{HMcode} as our halo model.

\begin{table}[]
\centering
\begin{tabular}{ccccc}
\toprule
\hline
\multicolumn{2}{c}{Parameter} &
\multicolumn{3}{c}
{Prior}\\
\hline
\multirow{6}{*}
{$\Lambda$CDM}
& $\Omega_{\rm b0} h^2$          & norm  & $0.0222$  & $0.0005$  \\
& $\Omega_{\rm c0} h^2$          & flat  & $0.001$   & $0.99$    \\
& $100\,\theta_{\rm MC}$          & flat  & $0.5$     & $10$      \\
& $\ln ( 10^{10} A_{\rm s} )$   & flat  & $1.61$    & $3.91$    \\
& $n_{\rm s}$                   & flat  & $0.8$     & $1.2$     \\
& $\tau_{\rm reio}$             & flat  & $0.01$    & $0.8$     \\
\hline
\multirow{1}{*}
{EPN special M}             & $M$                   & flat  & $-0.5$    & $0.5$       \\
\hline
\multirow{2}{*}
{EPN full special}             & $M$                   & flat  & $-0.5$    & $-10^{-8}$       \\
   & $\log_{10}c_m$   & flat  & $-1$      & $6$       \\
\hline
\bottomrule
\end{tabular}
\caption{The model parameters of $\Lambda$CDM and EPN considered in this work. We consider normal and flat prior distributions, indicated in the third column. In the former case, the fourth and fifth columns denote respectively the mean and width; in the latter case, they denote respectively the minimum and maximum.}
\label{table: prior}
\end{table}

A summary of cosmological parameters and their priors for different models are provided in Table~\ref{table: prior}. We assume the BBN prior on the baryon density $\Om{b0} h^2$ and a cut $50 < H_0 < 100$ on the Hubble constant $H_0$, translated here to $\theta_{\rm MC}$ \cite{Hu:1995en}. Following standard practice, the prior distributions for the latter as well as for all remaining parameters are chosen as flat. Additionally, we consider a single massive neutrino and two massless neutrinos in the calculations, fixing 
the sum of neutrino masses as $\sum m_\nu=0.06$ \text{eV}. The total matter density parameter at $\Omega_{\rm m0}$ thus includes the contributions of cold dark matter, baryons, and one neutrino species.

We make use of the \texttt{Cobaya} 
\cite{Torrado:2020dgo, Lewis:2002ah, Lewis:2013hha, 2005math......2099N} Markov Chain Monte Carlo (MCMC) sampler to generate the posterior distribution of the full set of cosmological parameters. For each dataset, two chains are deployed based on the early convergence test, and we consider the chains as converged when the stopping criteria of intra-chain and inter-chain $R-1 < 0.01$ are satisfied.\footnote{The parameter $R = \underset{\theta}{\max} \sqrt{ 1 + {\rm var}_{i} \big{(} {\rm mean}_{{\rm seg}_i} (\theta) \big{)} / {\rm mean}_{i} \big{(} {\rm var}_{{\rm seg}_i} ( \theta ) \big{)} }$ probes the consistency of parameter distributions $P(\theta)$ across chains for inter-chain $R$ or segments within a particular chain for intra-chain $R$.} We use 
\texttt{GetDist} \cite{Lewis:2019xzd} to perform statistical analyses of MCMC samples and to plot the posteriors.\vspace{0.3cm}

The  list of observational datasets and likelihoods used in this work is:
\begin{itemize}
\item 
{\bf Cosmic Microwave Background (CMB)} temperature and polarization anisotropy measurements from the \textit{Planck} satellite. This rich and well-analyzed dataset is the main basis of our analysis.\footnote{The use of the CMB dataset sets our work apart from \cite{Anagnostopoulos:2023pvi, Sudharani:2024qnn}. A comparison with these references would require the exclusion of the CMB dataset.} The CMB dataset used in this work
includes:  
(i) \textit{Planck} 2018 low multipole ($2 \le \ell\le 30$) temperature anisotropy power spectrum $C^{TT}_{\ell}$, reconstructed using the \texttt{Commander} likelihood; 
(ii) \textit{Planck} 2018 low multipole ($2 \le \ell\le 30$) large-scale E-mode polarization power spectrum $C^{EE}_{\ell}$, derived from the \texttt{Simall} likelihood; 
(iii) high multipole  power spectra of temperature and polarization anisotropies, $C^{TT}_{\ell}$, $C^{TE}_{\ell}$, $C^{EE}_{\ell}$, derived from the \texttt{NPIPE PR4 Planck CamSpec} likelihood \cite{Rosenberg:2022sdy}; 
(iv) lensing potential power spectrum, derived from NPIPE PR4 lensing reconstruction data \cite{Carron:2022eyg, Carron:2022eum}. We denote the combination of these likelihoods as ``{\bf CMB}''.

\item 
{\bf The Dark Energy Spectroscopic Instrument (DESI)} has measured the baryon acoustic oscillation (BAO) signal from galaxy clustering correlations using tracers from galaxies, quasars and the Ly$\alpha$ forest in the redshift range $0.1\le z\le 4.2$. We utilize 12 BAO measurements from the DESI Data Release 1 \cite{DESI:2024mwx, DESI:2024lzq, DESI:2024uvr}, which we denote here as ``{\bf DESI}''.

\item 
{\bf The PantheonPlusSH0ES supernova catalog}, containing the PantheonPlus sample \cite{Brout:2022vxf} (comprising 1701 light curves measured from 1550 Type Ia supernovae over the redshift range $0.001 < z < 2.26$), removing the data points in the redshift range $ z < 0.01$, with anchoring of SNIa standardized absolute magnitudes with SH0ES \cite{Riess:2021jrx}. We will refer to the combination of these supernova data as ``{\bf PPS}''.

\item
{\bf Galaxy Survey:} We also consider BAO, growth rate and $\sigma_8$ measurements by the 6-degree Field (6dF) Galaxy Survey \cite{Beutler:2012px, Ross:2014qpa} and the Sloan Digital Sky Survey (SDSS) Data Release 16 \cite{eBOSS:2019dcv}, collectively referred to as ``{\bf SDSS}'' here. This dataset is mutually exclusive of DESI BAO dataset.\footnote{Due to overlapping regions of observation between SDSS and DESI, the DESI group has cautioned against na\"{i}ve combinations of the two datasets \cite{DESI:2024mwx}. For this reason, we only consider SDSS and DESI data individually in our analysis.}
\end{itemize}

We also present two supplementary datasets in some of our plots that are excluded from the cosmological fit. These include an additional $f\sigma_8$ dataset (table 2 of \cite{Avila:2022xad}, with data entries from \cite{eBOSS:2020lta, Turnbull:2011ty, Achitouv:2016mbn, Beutler:2012px, Feix:2015dla, BOSS:2016wmc, BOSS:2013eso, Blake:2012pj, Nadathur:2019mct, BOSS:2013mwe, Wilson:2016ggz, eBOSS:2018yfg, Okumura:2015lvp}), SDSS DR7 LRG dataset \cite{Reid:2009xm} and eBOSS DR14 Ly-$\alpha$ dataset \cite{eBOSS:2018qyj} (presented in Fig.~\ref{fig: fsgma8-z}), and the weak lensing dataset from the Dark Energy Survey (DES) Y3 data release (presented in Fig.~\ref{fig: contour SDSS}).

\paragraph{Information Criteria IC \& Bayesian Evidence:} 
To compare between different datasets and models, we primarily utilize 6 probes of {\it Information Criteria (IC)}, defined in Appendix~\ref{sec: Information Criteria}. The traditional Bayesian evidence $B$, the {\it Deviance Information Criteria (DIC)} and the {\it Widely Applicable Information Criteria (WAIC)} \cite{10.5555/2567709.2502609} are for model comparison, while the Bayesian ratio $R$, the goodness of fit, and the suspiciousness \cite{DES:2020hen, Raveri:2019gdp, PhysRevD.100.023512} are employed for data tension detection. While probes such as DIC, WAIC and Bayesian ratio can be easily interpreted using the Jeffreys’ scale (see Table~\ref{table: Jeffrey’s scale}), the Goodness-of-Fit and Suspiciousness require further processing. As these two probes follow the BMD-dimensional $\chi$ distribution \cite{PhysRevD.100.023512}, we may convert them into the usual confidence level $\sigma$ value as ${\rm CDF}_1^{-1} \left( {\rm CDF}_{\rm BMD} \left( \sqrt{{\rm BMD} + 2z} \right) \right)$, where ${\rm CDF}_d$ is the cumulative distribution function of a $d$-dimensional $\chi$ distribution and $z$ stands for either the Goodness-of-Fit or Suspiciousness; see Appendix~\ref{sec: Information Criteria} for the definitions.

\section{Results}
\label{sec: Results}

This Section presents the results of our analysis. In Sec.~\ref{subsec: Parameter Constrain} we present cosmological constraints on the parameters $H_0$, $\Omega_{\rm m0}$, $S_8$ and $M$ obtained from the $\Lambda$CDM and EPN models at the background-modified level. In Sec.~\ref{subsec: Matter power spectrum and growth rate} we compare the matter power spectrum and growth rate obtained in $\Lambda$CDM and EPN special model with and without perturbations, paying particular attention to the effects of the mass parameter $c_m$. In Sec.~\ref{sec: Model comparison} we evaluate and compare the performance of all models according to the 
IC described in Sec.\ \ref{sec: Data and Analysis setting Methodology} and in Appendix~\ref{sec: Information Criteria}, discussing in particular the discrepancy between early- and late-time universe fits through the CMB TT power spectra and distance modulus reconstructed from the MCMC chains.

\subsection{Parameter constraints}
\label{subsec: Parameter Constrain}

\begin{table}
\centering
\scriptsize
\begin{tabular}{lccccc}
\toprule
\midrule
Model/Parameters  &CMB  &CMB + DESI &CMB + DESI + PPS &DESI &PPS \\
\midrule
$\mathbf{\Lambda CDM}$ &  &  &  &  &  \\
{$H_0$ [$\mathrm{km}\ \mathrm{s}^{-1}\ \mathrm{Mpc}^{-1}$]}
& $67.21\pm 0.46$
&$67.86\pm 0.37$ &$68.40\pm 0.34$
& $68.70\pm 0.79$ &$73.6\pm 1.0$\\
{$\Omega_{\rm{m0}}$}
&$0.3158\pm 0.0064$
&$0.3067\pm 0.0049$ &$0.2999\pm 0.0044$
& $0.295\pm 0.015$ &$0.332\pm 0.018$\\
{$\sigma_8$}  
& $0.8078\pm 0.0054$ &$0.8066\pm 0.0056$ &$0.8060\pm 0.0059$& $--$ &$--$\\
{$S_8$}  & $0.829\pm 0.011$&$0.8156\pm 0.0091$ &$0.8058\pm 0.0088$& $--$ &$--$ \\
\midrule
\textbf{EPN ${M=-1/3}$} &  &  &  &  &  \\
{$H_0$ [$\mathrm{km}\ \mathrm{s}^{-1}\ \mathrm{Mpc}^{-1}$]} 
& $71.91\pm 0.50$
&$71.66\pm 0.39$ &$71.52\pm 0.36$
&$72.36\pm 0.89$ &$73.6\pm 1.0$\\
{$\Omega_{\rm{m0}}$}  
& $0.2753\pm 0.0057$
&$0.2780\pm 0.0043$ &$0.2798\pm 0.0041$
& $0.298^{+0.013}_{-0.015}$ &$0.378\pm 0.019$\\
{$\sigma_8$} 
& $0.8344\pm 0.0055$&$0.8281\pm 0.0056$ &$0.8286\pm 0.0055 $& $--$ &$--$ \\
{$S_8$}  & $0.799\pm 0.011$ &$0.7971\pm 0.008$ &$0.8001\pm 0.0083$& $--$ &$--$\\
\midrule
\textbf{EPN ${M=-1/2}$} &  &  &  &  &  \\
{$H_0$ [$\mathrm{km}\ \mathrm{s}^{-1}\ \mathrm{Mpc}^{-1}$]}  
& $73.54\pm 0.53$
&$73.11\pm 0.39$ &$72.68\pm 0.37$
& $73.84\pm 0.87$ &$73.7\pm 1.0$\\
{$\Omega_{\rm{m0}}$}  
& $0.2635\pm 0.0056$
&$0.2681\pm 0.0041$ &$0.2728\pm 0.0040$
& $0.298\pm 0.014$ &$0.397\pm 0.019$\\
{$\sigma_8$}  & $0.8352\pm 0.0056$ &$0.8361\pm 0.0056$ &$0.8367\pm 0.0057$& $--$ &$--$\\
{$S_8$}  & $0.783\pm 0.011$&$0.7905\pm 0.0088$ &$0.7979\pm 0.0084$& $--$ &$--$ \\
\midrule
\textbf{EPN special M} &  &  &  &  &  \\
{$H_0$ [$\mathrm{km}\ \mathrm{s}^{-1}\ \mathrm{Mpc}^{-1}$]}  & 
$65.3^{+5.1}_{-4.3}$&$69.9\pm 1.2 $ &$69.82\pm 0.63$& $68.9^{+3.7}_{-2.4}$ &$73.5\pm 1.0$ \\
{$\Omega_{\rm{m0}}$}  & $0.340^{+0.034}_{-0.062}$&$0.2912\pm 0.0098$ &$0.2905\pm 0.0054$
& $0.294\pm 0.016$ &$0.315^{+0.058}_{-0.051}$ \\
{$\sigma_8$}  & $0.75^{+0.17}_{-0.35}$ &$0.8180\pm 0.0086$ &$ 0.8164\pm 0.0067$& $--$ &$--$\\
{$S_8$}  & $0.846^{+0.034}_{-0.044}$ &$0.806\pm 0.011$ &$0.8033\pm 0.0087$& $--$ &$--$\\
\midrule
{$M$}  & $0.12^{+0.32}_{-0.15}$
&$-0.164^{+0.11}_{-0.091}$ &$ -0.136^{+0.057}_{-0.051}$
& $-0.05\pm 0.23$ &$> -0.0924$ \\
\midrule
\bottomrule
\end{tabular}
\caption{Cosmological parameter constraints from different data combinations and priors, in the $\Lambda$CDM and three EPN special M models ($M=-1/3$, $M=-1/2$, and $M$ as fitting parameter). Results are quoted for the marginalized means and $68\%$ confidence intervals or lower limit.}
\label{table: results background}
\end{table}

\begin{table}
\centering
\scriptsize
\begin{tabular}{lcc}
\toprule
\midrule
Model/Parameters  &CMB  &CMB + SDSS \\
\midrule
$\mathbf{\Lambda CDM}$ &  &  \\
{$H_0$ [$\mathrm{km}\ \mathrm{s}^{-1}\ \mathrm{Mpc}^{-1}$]}
& $67.21\pm 0.46$
&$67.56\pm 0.36$ \\
{$\Omega_{\rm{m0}}$}
&$0.3157\pm 0.0065$
&$0.3109\pm 0.0049$ 
\\
{$\sigma_8$}  
& $0.8075\pm 0.0056$ &$0.8084\pm 0.0054$ \\
{$S_8$}  & $0.828\pm 0.011$&$0.8229\pm 0.0091$  \\
\midrule
{$\text{DIC}$}
& $5498.05 \pm 0.10$
&$5510.10\pm 0.40$ \\
{$\text{WAIC}$}
& $5498.95\pm0.07$
&$5511.04\pm0.24$ \\
{$-\ln B$}
&$5498.59\pm0.27$
&$5510.90\pm1.12$ \\
\midrule
\textbf{EPN ${M=-1/3}$} &  &   \\
{$H_0$ [$\mathrm{km}\ \mathrm{s}^{-1}\ \mathrm{Mpc}^{-1}$]} 
& $71.91\pm 0.50$
&$71.21\pm 0.40$ \\
{$\Omega_{\rm{m0}}$}  
& $0.2753\pm 0.0057$
&$0.2832\pm 0.0046$ \\
{$\sigma_8$} 
& $0.8344\pm 0.0055$&$0.8348\pm 0.0058$ \\
{$S_8$}  & $0.799\pm 0.011$ &$0.8112\pm 0.0095$\\
\midrule
{$\Delta \text{DIC}$}
& $0.42\pm 0.15$
&$ 4.31\pm 0.37$ \\
{$\Delta \text{WAIC}$}
& $1.09\pm0.77$
& $4.78\pm0.37$ \\
{$\Delta -\ln B$}
&$-0.66\pm2.69$
&$5.03\pm0.58$ \\
\midrule
\textbf{EPN very special} &  &   \\
{$H_0$ [$\mathrm{km}\ \mathrm{s}^{-1}\ \mathrm{Mpc}^{-1}$]}  
& $72.21\pm 0.56$
&$71.29\pm 0.41$ \\
{$\Omega_{\rm{m0}}$}  
& $0.2719\pm 0.0061$
&$0.2824\pm 0.0047 $ \\
{$\sigma_8$}  & $0.8499\pm 0.0059$ &$0.8494\pm 0.0060$ \\
{$S_8$}  & $0.809\pm 0.011$&$0.8241\pm 0.0091$  \\
{$\log_{10}c_m$}  & $ 4.42^{+0.66}_{-1.1}$&$ 4.83^{+1.1}_{-0.49}$  \\
\midrule
{$\Delta \text{DIC}$}
& $-0.15 \pm 0.26$
&$5.28\pm 0.48$ \\
{$\Delta \text{WAIC}$}
& $0.13\pm 0.91$
&$5.31\pm 0.24$ \\
{$\Delta -\ln B$}
&$-0.31\pm 2.02$
&$4.19\pm 1.10$ \\
\midrule
\textbf{EPN full special} &  &   \\
{$H_0$ [$\mathrm{km}\ \mathrm{s}^{-1}\ \mathrm{Mpc}^{-1}$]}  
&$71.41\pm 0.81$
&$70.16^{+0.46}_{-0.56}$ \\
{$\Omega_{\rm{m0}}$}  
& $0.2777\pm 0.0075$
&$0.2898\pm 0.0053$ \\
{$\sigma_8$}  & $0.8424\pm 0.0082$ &$0.8365\pm 0.0071$ \\
{$S_8$}  & $0.810\pm 0.011$&$0.8221\pm 0.0092$  \\
{$M$}  & $-0.260^{+0.051}_{-0.076}$&$ -0.202^{+0.046}_{-0.024}$  \\
{$\log_{10}c_m$}  & $4.65^{+1.0}_{-0.77}$&$ > 5.08$  \\
\midrule
{$\Delta \text{DIC}$}
& $-0.40 \pm 0.20 $
&$2.22\pm 0.41 $ \\
{$\Delta \text{WAIC}$}
& $-0.23\pm 0.14$
&$2.95\pm 0.27 $ \\
{$\Delta -\ln B$}
&$-0.72\pm 2.32 $
&$2.54\pm 1.41 $ \\
\midrule
\bottomrule
\end{tabular}
\caption{Cosmological parameter constraints from different data combination and priors, in the $\Lambda$CDM, the special M with $M=-1/3$ (without dark energy perturbations), the very special and the full special models. Results are quoted for the marginalized means and $68\%$ confidence intervals or lower limit.
}
\label{table: parameter constrain SDSS}
\end{table}

\begin{figure}
\centering
\includegraphics[width=0.5\linewidth]{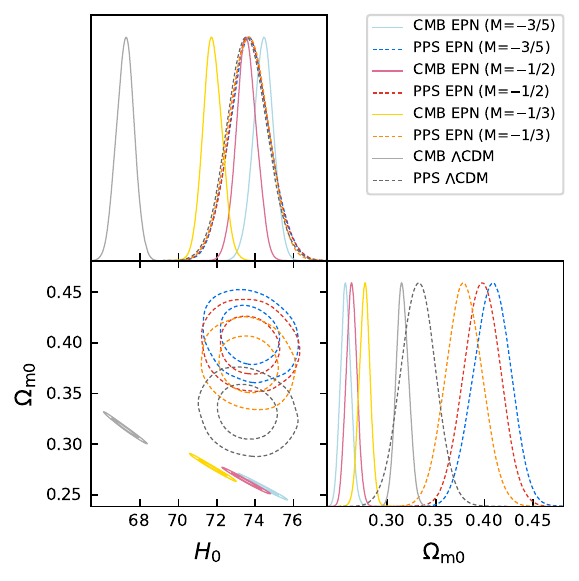}
\caption{Marginalized posterior constraints on $H_0$ and $\Om{m0}$ using CMB and PPS data in $\Lambda$CDM and three EPN special M models. Contours indicate 68\% and 95\% confidence level intervals.}
\label{fig: contour CMB vs SN}
\end{figure}

\begin{figure}
\centering
\includegraphics[width=0.6\linewidth]{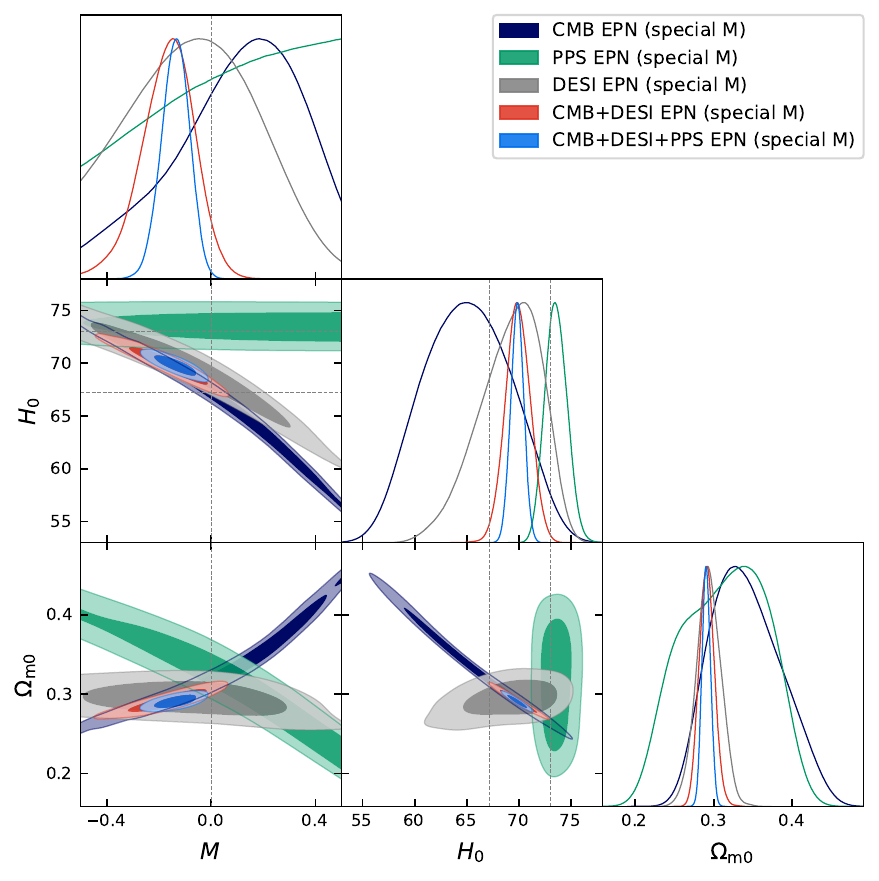}
\caption{Marginalized posterior constraints on $M$, $H_0$ and $\Omega_{\rm m0}$ using different (combinations of) datasets in the EPN special M model. Contours indicate 68\% and 95\% confidence level intervals. Dashed lines indicate best fit values for $\Lambda$CDM, i.e.\ with $M=0$. Although the CMB, DESI and PPS datasets do not individually constrain the parameter $M$ well, the combined dataset CMB + DESI + PPS favors $M<0$ at the $\gtrsim 2\sigma$ confidence level ($\gtrsim1\sigma$ for CMB + DESI).} 
\label{fig: contour_General M}
\end{figure}

Table~\ref{table: results background} summarizes the parameter constraints obtained from the data fits of $\Lambda$CDM, two EPN models with fixed $M$ (the special M model with $M=-1/3$ and the model $M=-1/2$ motivated by the considerations of Sec.\ \ref{subsec: EPN nonminimal}) as well as the case with free $M$. Recall that we begin by focusing on the background-modified level, i.e.\ ignoring perturbations of the EPN field, but including metric and matter perturbations; see Sec.\ \ref{sec: Data and Analysis setting Methodology} for details on our set-up.

Remarkably, we see that the result for $H_0$ in the EPN special M model with $M=-1/3$ is $71.91\pm 0.50\ \mathrm{km}\ \mathrm{s}^{-1}\ \mathrm{Mpc}^{-1}$ when derived from CMB data and $73.6\pm 1.0\ \mathrm{km}\ \mathrm{s}^{-1}\ \mathrm{Mpc}^{-1}$ from supernova PPS data. The Hubble tension in this set-up is therefore significantly reduced down to $\sim1.52\sigma$ (in comparison with $\sim5.81\sigma$ in $\Lambda$CDM). Furthermore, when leaving $M$ as a fitting parameter, we find that the more negative $M$ is, the higher $H_0$ derived from the CMB dataset becomes, until it matches the local $H_0$ value of PPS at $M \simeq -1/2$ (cf.\ Table \ref{table: prior}). This is fully consistent with what is found in phenomenological models, such as the CPL parameterization, and other phantom dark energy models: at the background level, a rapid increase in the dark energy density at late times can alone result in an increase of $H_0$, while maintaining the distance to the last scattering surface in order to match the acoustic peak scale given by CMB data. See e.g.~\cite{Knox:2019rjx,DiValentino:2021izs}.

On the other hand, along with the decrease in the Hubble tension, a tension in the measurement of $\Omega_{\rm m0}$ from CMB and PPS datasets arises; see Fig.~\ref{fig: contour CMB vs SN}. This behavior is again consistent with other phantom-like models (see e.g.\ \cite{deFelice:2017paw,Alestas:2020mvb}). The effect may be understood as follows. Although phantom dark energy increases $H_0$, as mentioned above, it also dims type-Ia supernovae more than $\Lambda$CDM does. The only way to compensate for this dimming, given the local $H_0$ measurements of SH0ES, is to increase $\Omega_{\rm m0}$, which in turn reduces the amount of dark energy. This is reflected by the comparatively large $\Omega_{\rm m0}$ deduced in EPN from the PPS dataset; cf.\ Table~\ref{table: results background}.

We also discover that the result for $S_8$ in the very special EPN model inferred from the CMB dataset is in mild tension with that deduced from the SDSS dataset, as shown in Table~\ref{table: parameter constrain SDSS}. This tension is even more evident in Fig.~\ref{fig: matter power_spectrum}, which shows that EPN models consistently predict stronger matter perturbation spectra compared with $\Lambda$CDM at and below the BAO acoustic scale.\footnote{As each dataset measures matter spectrum power at a different redshift, Fig.~\ref{fig: matter power_spectrum} pulls all data to the present day via the transfer function. This makes the data points model-dependent. Despite this caveat, we can still conclude that EPN models do introduce tension between CMB and SDSS, given that the models swing from weakly favored over $\Lambda$CDM for the CMB dataset to moderately disfavored for the CMB + SDSS dataset, as shown in Table~\ref{table: parameter constrain SDSS}.} However, this is not necessarily a drawback considering the derived value $S_8 = 0.776\pm 0.017$ for $\Lambda$CDM from DES Y3 \cite{DES:2020hen}, as is manifest in Fig.~\ref{fig: S8 distribution}. This behavior is again not specific to EPN, as the same behavior happens for the CPL phenomenological model with a matching cosmographic equation of state \cite{Alestas:2020mvb}. 
Quantifying more precisely by how much EPN performs better with the DES weak lensing dataset is left to further studies as including these data within our current  MCMC analysis proves to be computationally demanding and beyond the direct remit of this work.\footnote{The $S_8$ probe, designed to eliminate the misalignment between $f\sigma_8$ and $\sigma_8$ via the theoretically motivated formula $f\sim \Om{M}^{0.55}$, no longer describe the matter perturbation history correctly in EPN models, as shown in Fig.~\ref{fig: fsgma8-z}. It is therefore difficult to interpret the $S_8$ values presented in Table~\ref{table: parameter constrain SDSS}, in particular when compared with the derived value of $S_8$ from the DES dataset. However, we do treat SDSS data as likelihoods over $f\sigma_8$. All the tension probes and IC values presented in Table~\ref{table: parameter constrain SDSS} thus correctly evaluate the SDSS dataset, including the comparison between different models.}

More interestingly, when allowing for a variable EPN parameter $M$ in the analysis, the CMB + DESI + PPS combination constrains $M$ to $-0.136^{+0.057}_{-0.051}$, i.e.\ a clear departure from $\Lambda$CDM ($M=0$) at the $\gtrsim$ $2\sigma$ level, while for the CMB + DESI combination the result is $M=-0.164^{+0.11}_{-0.091}$, which is less tight but still $\gtrsim$ $1\sigma$.\footnote{This is a looser constraint than the one obtained in \cite{DeFelice:2020sdq} in the context of GP theory; the reason is likely due to the omission in that reference of SNIa anchoring, which is known to be detrimental to phantom-like models, cf.\ Fig.~\ref{fig: chi_sn}. Note incidentally that the parameter $s$ in \cite{DeFelice:2020sdq} is equivalent to $-M$ here.} Fig.~\ref{fig: contour_General M} displays the constraints on $M$, $H_0$ and $\Om{m0}$ and their posterior distributions. See also Fig.\ \ref{fig: contour_CMB+DESI+PPS 6 parameters} in Appendix \ref{sec:appendix full constraint plot} for the complete constraint plot of cosmological parameters from the CMB + DESI + PPS dataset.


\subsection{Matter power spectrum and structure growth}
\label{subsec: Matter power spectrum and growth rate}

\begin{figure}
\centering
\subfigure[\label{fig: S8 distribution}]{\includegraphics[width = 0.55 \textwidth]{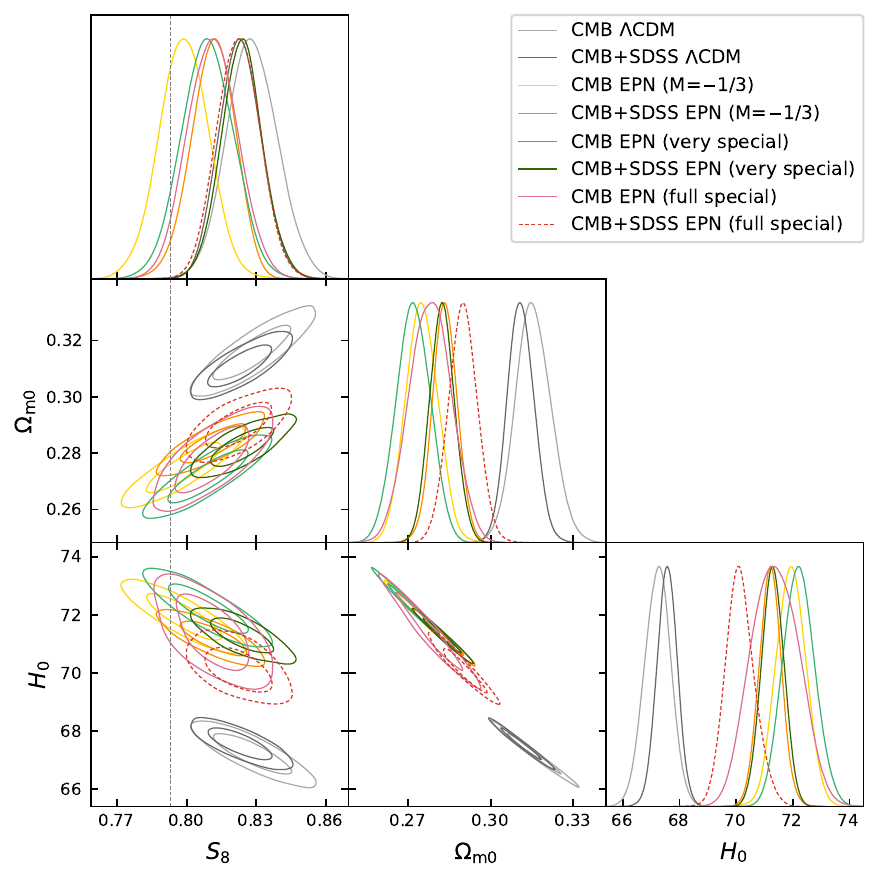}}~
\subfigure[\label{fig: m distribution}]{\includegraphics[width = 0.38 \textwidth]{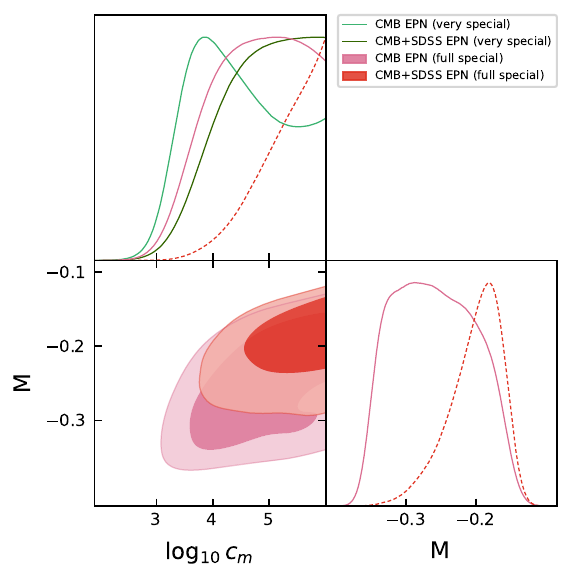}}
\caption{\emph{Left panel}: 68\% and 95\% marginalized posterior constraints on $S_8$, $\Omega_{\rm m0}$ and $H_0$ using CMB, CMB+SDSS datasets in $\Lambda$CDM, the EPN special M ($M=-1/3$), very special and full special models. The vertical dashed line in $S_8$ corresponds to the best fit + 1$\sigma$ value $S_8=0.776\pm0.017$ derived for $\Lambda$CDM from DES Y3. \emph{Right panel}: marginalized posterior constraints on $M$ and the Proca mass parameter $c_m$ in the EPN very special and full special models from the same two datasets.
}
\label{fig: contour SDSS}
\end{figure}

\begin{figure}
\centering
\includegraphics[width=0.5\linewidth]{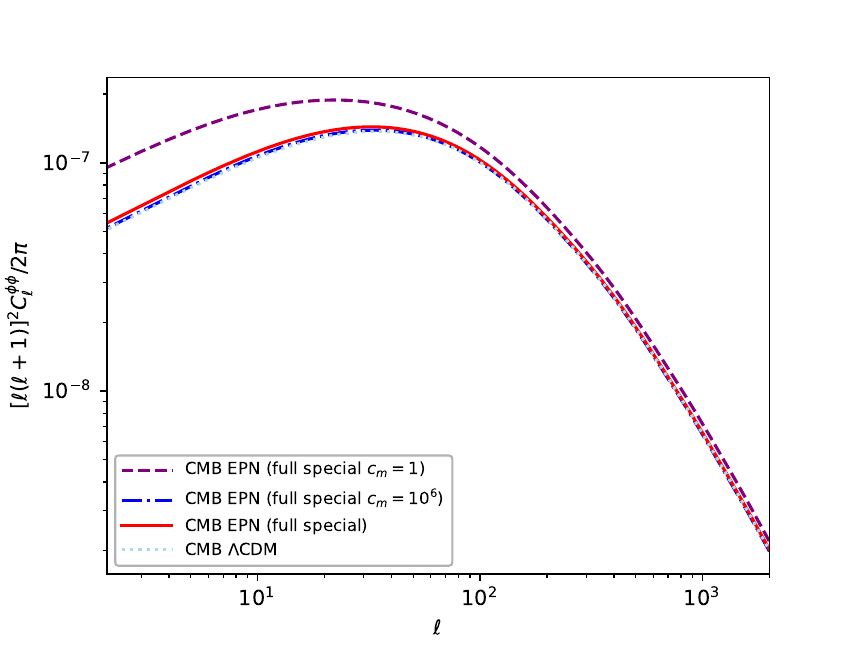}
\caption{The lensing potential of the best-fit $\Lambda$CDM model and the EPN full special model given the best-fit parameters for the CMB dataset, with different values of the Proca mass parameter $c_m$, including the best-fit value $c_m=10^{4.48}$ ({\color{red}red} curve). The strong enhancement for $c_m = \mathcal{O}(1)$ values is manifest.}
\label{fig: cmb lensing}
\end{figure}

\begin{figure}
\centering
\subfigure[Fitted against CMB.]{
\begin{tikzpicture}[      
        every node/.style={anchor=south west,inner sep=0pt},
        x=0.43mm, y=0.43mm,
      ]   
     \node (fig2) at (0,0)
       {\includegraphics[width =0.43 \textwidth]{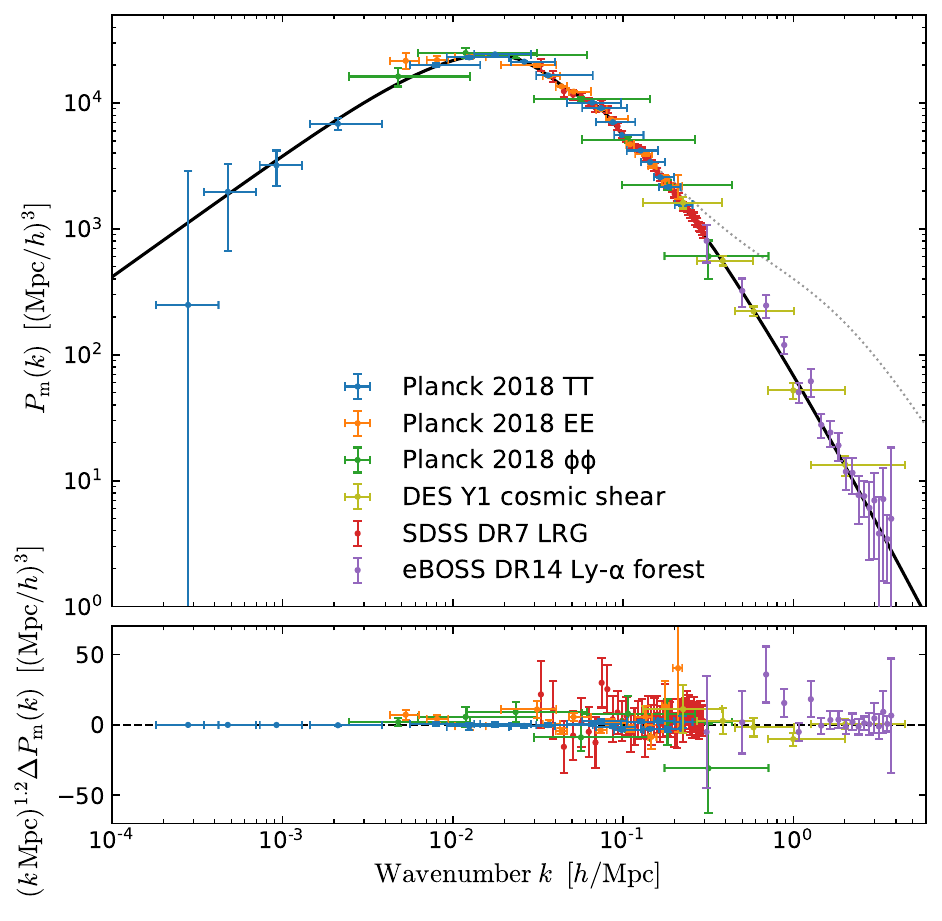}};
     \node (fig1) at (-2.8,10.7)
       {\includegraphics[width = 0.47945\textwidth]{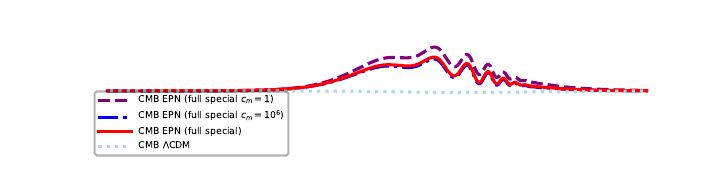}};
\end{tikzpicture}
}
\subfigure[Fitted against CMB + SDSS.]{
\begin{tikzpicture}[      
        every node/.style={anchor=south west,inner sep=0pt},
        x=0.43mm, y=0.43mm,
      ]   
     \node (fig2) at (0,0)
       {\includegraphics[width =0.43 \textwidth]{tegfig.pdf}};
     \node (fig1) at (-2.8,10.7)
       {\includegraphics[width = 0.47945\textwidth]{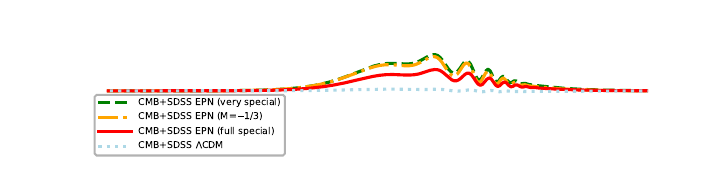}};
\end{tikzpicture}
}
\caption{\label{fig: matter power_spectrum}
Best-fit matter power spectrum (\emph{top panels}) and the residue (\emph{bottom}) against the best-fit $\Lambda$CDM spectrum. In the top panels (which are identical) solid curves represent the linear power spectrum while dotted curves represent the non-linear power spectrum (data taken from \cite{Chabanier:2019eai}). \textit{Bottom-left:} Residue plot for the EPN full special model from the CMB dataset, for fixed and varying $c_m$ values (same color-coding as Fig.\ \ref{fig: cmb lensing}). \textit{Bottom-right:} Residue plot for the EPN very special, full special, and the background only $M=-1/3$ models, fitted against the CMB+SDSS dataset. There is an evident discrepancy between $\Lambda$CDM and the EPN models, in particular at and below the BAO scale, with the EPN special background-only model slightly closer to $\Lambda$CDM than the very special model. There is a modest power suppression upon including SDSS data.
}
\end{figure}

\begin{figure}
\centering
\subfigure[Fitted against CMB.]{\includegraphics[width = 0.49 \textwidth]{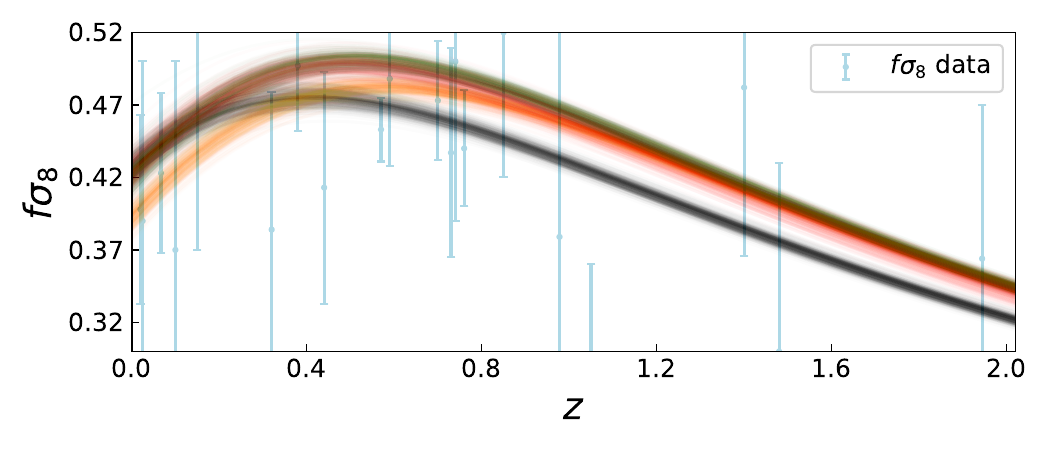}}~
\subfigure[Fitted against CMB + SDSS.]{\includegraphics[width = 0.49 \textwidth]{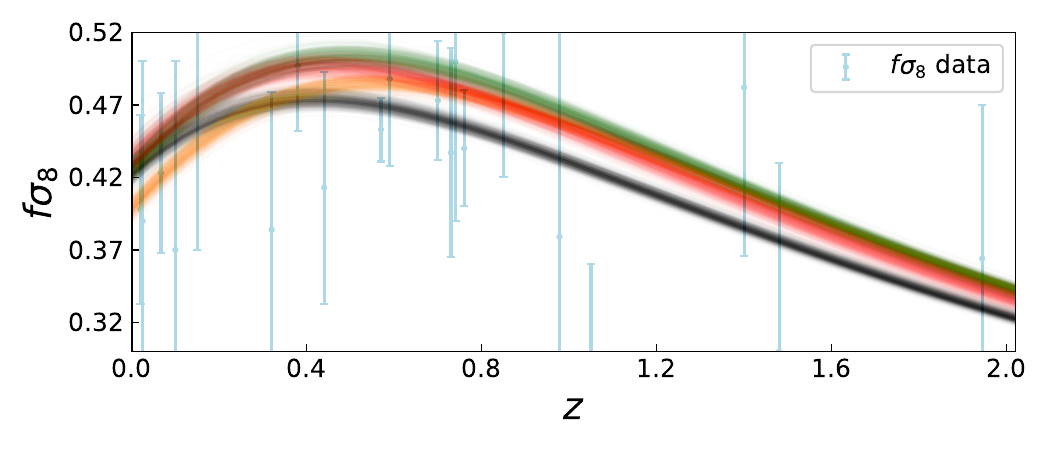}}
\caption{Structure growth rate $f\sigma_8$ as a function of redshift $z$. Curves show the fits against CMB (\emph{left panel}) and against CMB+SDSS (\emph{right}) datasets in $\Lambda$CDM (black), EPN special M model with $M=-1/3$ ({\color{Dandelion}yellow}), EPN very special model ({\color{ForestGreen}green}) and EPN full special model ({\color{red}red}). Data points were taken from Table~2 of \cite{Avila:2022xad}.}
\label{fig: fsgma8-z}
\end{figure}

Next we investigate the impact of perturbations in the data fits, considering the model realizations given by the EPN very special and full special models, cf.\ Sec.\ \ref{subsec: Evolution of the perturbation in the minimally coupled EPN model}. We focus in particular on the analysis of the structure growth history as inferred from the CMB and SDSS datasets. 

Fig.~\ref{fig: contour SDSS} shows the fitting contours obtained from different combinations of datasets as well as the constraints on the Proca mass parameter $c_m$. A parametrically large\footnote{By parametrically large, we mean a mass $m$ two or three orders of magnitude larger than the scale of dark energy, which is hence still well within the acceptable range. Given that the initial prior is flat and enables the code to explore much larger value of the mass, it is very encouraging to note that the favored value converges towards values of the mass which are within the theoretically reasonable range.} value of $c_m$ is strongly preferred for both EPN model realizations. As demonstrated by Fig.~\ref{fig: cmb lensing}, such a preference is induced by the foreground effect of the CMB, in particular the late-time integrated Sachs-Wolfe and the lensing effects as well as the matter perturbation growth history, since the EPN model introduces an enhancement of the effective Newton constant, $G_{\rm eff}$, altering the low-$\ell$ CMB spectrum and $f\sigma_8$;\footnote{Foreground effects may also be enhanced by anisotropic stress. However, we have checked that the EPN special model does not introduce anisotropic stress, in full analogy with GP theory \cite{DeFelice:2016uil,deFelice:2017paw}; see Appendix \ref{sec: CAMB}.} see Appendix \ref{sec: CAMB} for details. The excess spectrum power is due to the EPN scalar contribution to the scalar potential, which can be suppressed by taking the decoupling limit, hence the preference for parametrically large values of $c_m$ corresponding to $m\sim \mathcal{O}\left(10^{2}H_0\right)$, where $H_0$ is the Hubble parameter today, so this is still within the acceptable range for the mass. It is also worth commenting on the upper confidence intervals on $c_m$ quoted in Table \ref{table: parameter constrain SDSS}. In view of the irregular posterior distributions, cf.\ Fig.\ \ref{fig: contour SDSS}(b), this should not be seen as an upper bound but rather as indicating that excessively large mass values are less preferred. This is likely to compensate for the excessive `phantomness' of the EPN field at the background level. The results of our analysis therefore do not confirm the claim made in \cite{Savaliya:2025cct} on the existence of an upper bound on the mass of the Proca field from cosmological observations. Rather our analysis indicate no clear sign of an upper bound.

The $f\sigma_8$ measurements from $\Lambda$CDM and the EPN special model with and without dark energy perturbations are presented in Fig.~\ref{fig: fsgma8-z}. The tendency of a greater $f\sigma_8$ before the dark energy-dark matter transition, and a lower value afterwards, is consistent with the background evolution history, i.e.\ a phantom-like dark energy fluid leading to a greater $\Omega_\M$ before the transition and a lower $\Omega_\M$ afterwards.

One might naively have expected a sizable alteration of the growth rate at low redshifts from the modification of the matter perturbation equations by the EPN terms, cf.\ Eqs.~\eqref{eq:coefsomegaspec}. However, a careful derivation shows that these modifications are actually $\mathcal{O}(\Om{EPN}^2)$, implying a significant effect on the growth history only at very low redshifts and therefore essentially unconstrained by current observations. Curiously, the modification brings the $f\sigma_8$ value back to that of $\Lambda$CDM at around the present time, regardless of the value of the background parameter $M$. Whether this is a mere coincidence or a signal of hidden structures within EPN theory remains an open question.

Confirming expectations, our results highlight the negligible impact of taking into account versus ignoring dark energy perturbations in the data analysis upon considering the decoupling regime of parametrically large Proca masses. The naive reason why this had to be the case is that EPN is essentially a theory of dark energy with suppressed early-time modifications to cosmology. However, it is worth noticing that perturbations do lead to significant constraints on model parameters in the context of GP theory \cite{DeFelice:2020sdq}. In this regard, the EPN special model may be seen as a particularly motivated theory of dark energy, addressing the late-time cosmic acceleration puzzle while leaving plenty of parameter space left to address cosmological tensions.


\subsection{Model comparison}
\label{sec: Model comparison}

\begin{table}
\centering
\scriptsize
\begin{tabular}{lccccc}
\toprule
\midrule
Model/Parameters  &CMB &CMB + DESI &CMB + DESI + PPS&DESI&PPS  \\
\midrule
$\mathbf{\Lambda CDM}$ &  &  &  &  &  \\
{DIC}  & $5498.05 \pm 0.10$ &$5506.94\pm 0.16$&$6251.24\pm0.34$& $8.39\pm0.06$ &$728.04\pm0.002$\\
{WAIC}  & $5498.95\pm0.07$ &$5508.29\pm 0.68$&$6252.01\pm0.19$& $8.41\pm0.09$ &$728.08\pm0.03$\\
{$-\ln B$}  & $5498.59\pm0.27$ &$5509.75\pm 2.72$&$6251.25\pm0.14$& $8.85\pm0.10$ &$728.37\pm0.14$\\
\midrule
{Tension against}  & & & & CMB &CMB + DESI \\
\midrule
{$- \ln{R}$}  & & & & $2.31\pm 2.93$& $13.12\pm 2.92$ \\
{Goodness of Fit}  & & & & $2.53\pm 0.56\sigma$& $5.65\pm 0.30\sigma$\\
{Suspiciousness}  & & & & $2.04\pm 0.47\sigma$& $5.52\pm0.27\sigma$ \\
\midrule
\textbf{EPN ${M=-1/3}$} &  &  &  &  &  \\
{$\Delta \text{DIC}$}  
&$0.42\pm0.15$
&$0.19\pm0.33$ &$1.6\pm 0.42$
&$0.51\pm0.07$ 
&$0.45\pm0.07$\\
{$\Delta \text{WAIC}$}  &$1.09\pm0.76$
&$-0.55\pm0.91$ &$2.2\pm 0.22$
&$0.51\pm0.16$ 
&$0.43\pm0.15$\\
{$\Delta -\ln B$}  
& $-0.66 \pm 2.68$
&$-2.0\pm3.24$&$2.24\pm0.22$
& $0.28\pm0.30$ &$0.79\pm0.80$ \\
\midrule
{Tension against} & & & & CMB &CMB + DESI\\
\midrule
{$\Delta - \ln{R}$}  & & & & $-2.41\pm 3.11$& $3.83\pm 3.22$ \\
{$\Delta$ Goodness of Fit}  & & & & $-1.23\pm 0.66 \sigma$& $0.70\pm 0.40\sigma$ \\
{$\Delta$ Suspiciousness}  & & & & $-1.01\pm 0.49 \sigma$& $0.60\pm 0.35\sigma$\\
\midrule
\textbf{EPN $M=-1/2$} &  &  &  &  &  \\
{$\Delta \text{DIC}$}  
& $1.31\pm0.23$
&$2.80\pm 0.20$&$11.85\pm0.42$
& $1.11\pm0.15$ &$0.76\pm0.02$\\
{$\Delta \text{WAIC}$}  
& $1.56\pm0.31$
&$2.17\pm 0.87$&$11.94\pm0.26$
& $1.11\pm0.24$ &$0.75\pm0.04$\\
{$\Delta -\ln B$}  & $1.31\pm1.09$
&$ 0.28 \pm3.12$ &$13.51\pm3.01$
& $1.30\pm0.87$ &$0.94\pm 0.19$\\
\midrule
{Tension against}  & & & & CMB &CMB + DESI \\
\midrule
{$ \Delta- \ln{R}$}  & & & & $-2.68\pm 3.40$& $12.64\pm 4.21$\\
{$\Delta$ Goodness of Fit}  & & & & $-0.42\pm 0.57\sigma$& $1.52\pm 0.34\sigma$\\
{$\Delta$ Suspiciousness}  & & & & $-0.29\pm 0.49\sigma$& $1.51\pm 0.32 \sigma$ \\
\midrule
\textbf{EPN special M} &  &  &  &  &  \\
{$\Delta \text{DIC}$}  
& $0.53\pm0.19$
&$-0.35\pm0.38$&$-3.2\pm0.57$
& $0.61\pm0.07$ &$0.18\pm0.09$\\
{$\Delta \text{WAIC}$} 
& $0.45\pm0.11$
&$-0.62\pm 0.79$&$-2.4\pm0.41$
& $0.46\pm0.10$ &$-0.03\pm0.12$\\
{$\Delta -\ln B$}  
& $0.94\pm2.27$
&$-2.39\pm 3.91$&$-2.99\pm1.70$
& $0.82\pm0.96$ &$-0.18\pm0.30$\\
\midrule
{Tension against}  &  & &  & CMB &CMB + DESI\\
\midrule
{$\Delta- \ln{R}$}  & & & & $-4.16\pm4.28$& $-0.46\pm 3.99$ \\
{$\Delta$ Goodness of Fit}  & & & & $-0.64\pm 0.58\sigma$& $-0.18\pm 0.40\sigma$\\
{$\Delta$ Suspiciousness}  & & & & $-0.87\pm 0.47\sigma$& $-0.29\pm 0.39 \sigma$\\
\midrule
\bottomrule
\end{tabular}
\caption{Analysis information criteria, Goodness of Fit, and Suspiciousness from several data combinations, in $\Lambda$CDM and EPN special M models with $M=-1/3$, $M=-1/2$ and $M$ as fitting parameter.
}
\label{table: analysis information criteria}
\end{table}

\begin{figure}
\centering
\subfigure[Fitted against CMB + DESI.]{
\begin{tikzpicture}[      
        every node/.style={anchor=south west,inner sep=0pt},
        x=0.45mm, y=0.45mm,
      ]   
     \node (fig2) at (0,0)
       {\includegraphics[width = 0.45\textwidth]{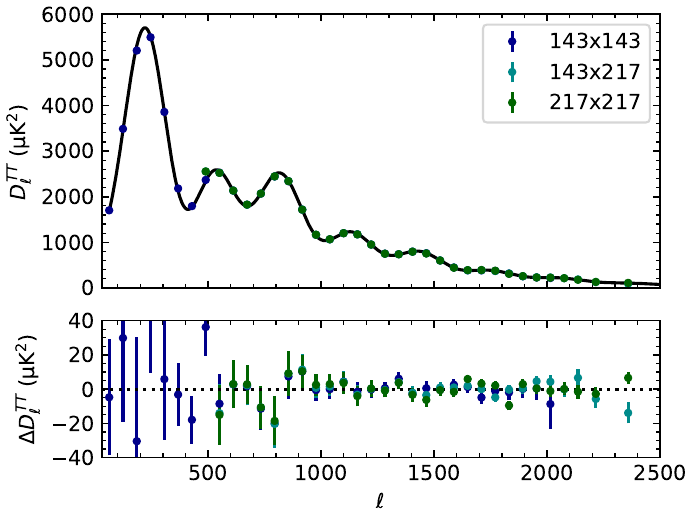}};  
     \node (fig1) at (2.7,9.1)
       {\includegraphics[width = 0.468\textwidth]{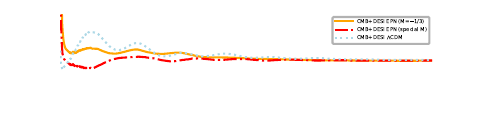}};
\end{tikzpicture}
}
\subfigure[Fitted against CMB + DESI + PPS.]{
\begin{tikzpicture}[      
        every node/.style={anchor=south west,inner sep=0pt},
        x=0.45mm, y=0.45mm,
      ]   
     \node (fig2) at (0,0)
       {\includegraphics[width = 0.45\textwidth]{npipe_TT_spec_residual_TTTEEE_fgsub.pdf}};  
     \node (fig1) at (2.7,9.1)
       {\includegraphics[width = 0.468\textwidth]{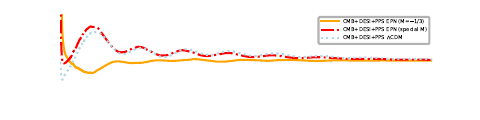}};
\end{tikzpicture}
}
\caption{\label{fig: High-l Comspec TTTEEE power_spectrum}
The best-fit CMB TT power spectrum (\emph{top panels}) and the residue (\emph{bottom}) against best-fit $\Lambda$CDM model TT power spectrum from the CMB dataset. The top panels are identical. Data points were taken from the \emph{Planck} CamSpec PR4 data release (from Fig.~6 of \cite{Rosenberg:2022sdy}). The residue plots correspond to the fits against CMB+DESI and CMB+DESI+PPS datasets, respectively. There are evident discrepancies between the fits against CMB data (black dotted line) and both against CMB+DESI and CMB+DESI+PPS datasets in $\Lambda$CDM. This tension is alleviated in EPN theory, improving compatibility between CMB and DESI measurements.
}
\end{figure}

\begin{figure}
\centering
\subfigure[Fitted against CMB.]{
\begin{tikzpicture}[      
        every node/.style={anchor= west,inner sep=0pt},
        x=0.43\textwidth, y=0.43\textwidth,
      ]   
     \node (fig2) at (0,0)
       {\includegraphics[width = 0.43 \textwidth]{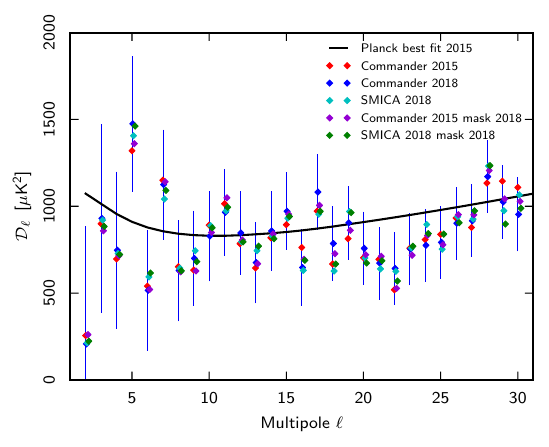}};
     \node (fig1) at (-0.008,0.035)
       {\includegraphics[width = 0.4687 \textwidth]{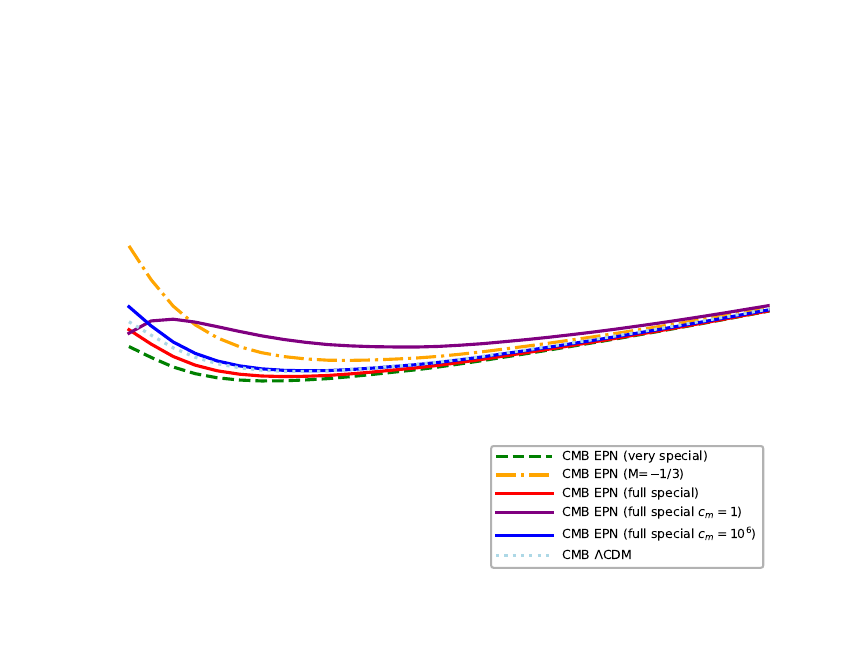}};
\end{tikzpicture}
}
\subfigure[Fitted against CMB + SDSS.]{
\begin{tikzpicture}[      
        every node/.style={anchor= west,inner sep=0pt},
        x=0.43\textwidth, y=0.43\textwidth,
      ]   
     \node (fig2) at (0,0)
       {\includegraphics[width = 0.43 \textwidth]{Commander_final.pdf}};
     \node (fig1) at (-0.008,0.035)
       {\includegraphics[width = 0.4687 \textwidth]{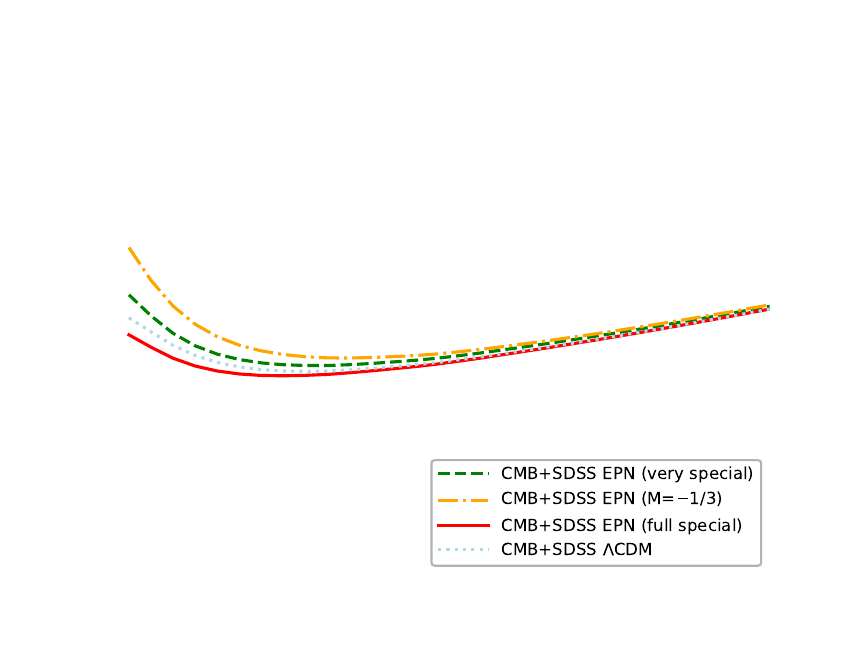}};
\end{tikzpicture}
}
\caption{\label{fig: Low-l Commander power_spectrum}
Same as Fig.\ \ref{fig: High-l Comspec TTTEEE power_spectrum} for CMB TT low-$\ell$ power spectra. Also shown are data of various maps and masks from the \emph{Planck}-18 data release (from Fig.~2 of \cite{Planck:2019nip}). Black solid line is the best-fit $\Lambda$CDM model prediction from \emph{Planck}-15. The left panel further highlights the effect of the EPN parameter $c_m$ on the low-$\ell$ spectrum, displaying results either with adjustable $c_m$ ({\color{ForestGreen}green} and {\color{red}red} curves, very special and full special models) or with fixed $c_m$ ({\color{violet}purple} and {\color{blue}blue} curves, full special model); for the latter, other parameters are chosen to match the best-fit values of the full special model.
}
\end{figure}

\begin{figure}
\centering
\subfigure[\label{fig: DL-z-CMB}Fitted against CMB]{\includegraphics[width = 0.49 \textwidth]{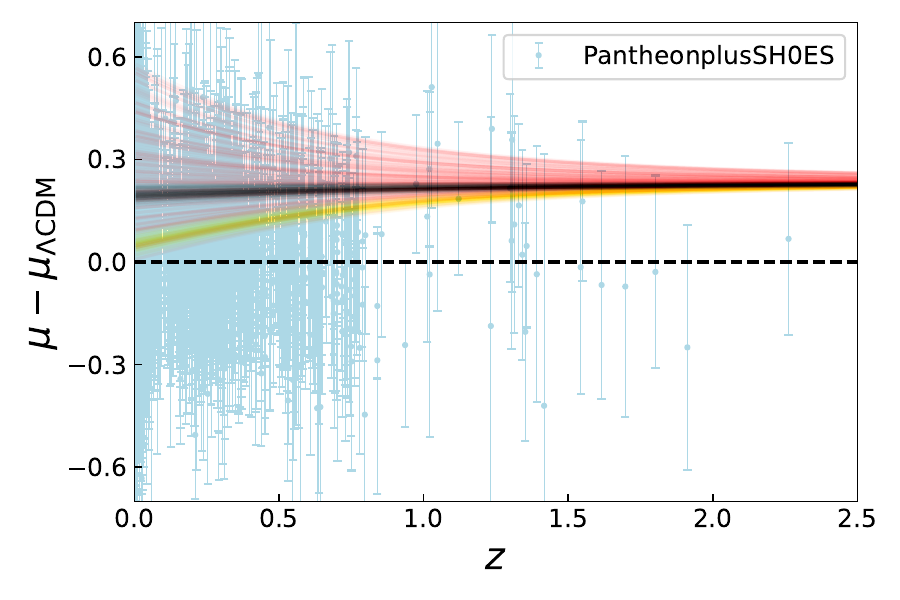}}~
\subfigure[\label{fig: DL-z-CMB+DESI}Fitted against CMB + DESI]{\includegraphics[width = 0.49 \textwidth]{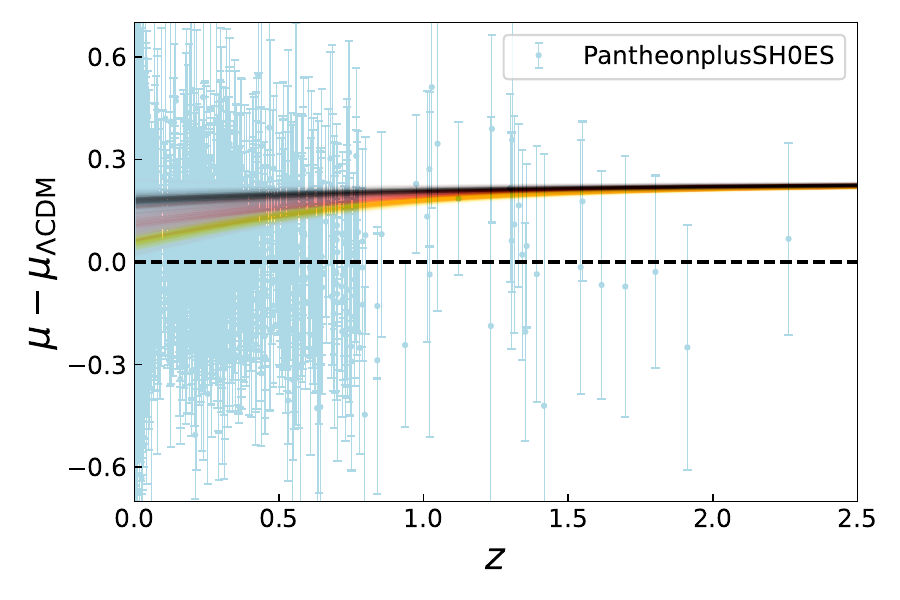}}
\subfigure[\label{fig: DL-z-CMB+DESI+PPS}Fitted against CMB + DESI + PPS]{\includegraphics[width = 0.49 \textwidth]{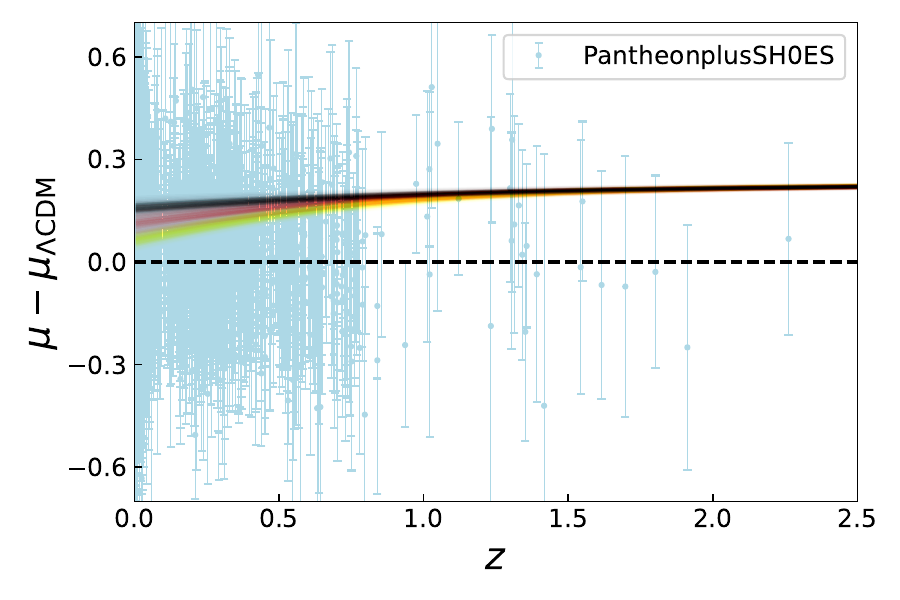}}~
\subfigure[\label{fig: chi_sn}$\chi^2_{SN}$ posterior distributions]{\includegraphics[width = 0.49 \textwidth]{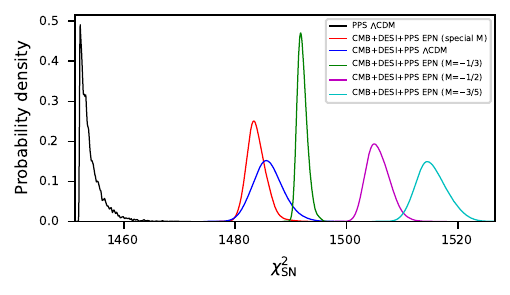}}
\caption{
\emph{Panels (a), (b), (c):} Reconstructed distance modulus fitted against CMB, CMB+DESI and CMB+DESI+PPS datasets in $\Lambda$CDM (black), EPN special M model with $M=-1/3$ ({\color{Dandelion}yellow}) and EPN special M model with adjustable $M$ ({\color{red}red}). The black dashed line in each figure represents the best-fit $\Lambda$CDM model from PPS data. \emph{Panel (d):} The $\chi^2_{\rm SN}$ posterior distributions from PPS in $\Lambda$CDM and CMB+DESI+PPS in $\Lambda$CDM and four realizations of the EPN special M model ($M=-1/3,-1/2,-3/5$ and fitting $M$).
}
\label{fig: DL-z}
\end{figure}

Table~\ref{table: analysis information criteria} shows the results of various IC measures, Goodness of Fit and Suspiciousness obtained for each model and dataset combination. Following the Jeffreys scale (cf.\ Table~\ref{table: Jeffrey’s scale}), we find the special EPN model with $M=-1/2$ to be ruled out by the CMB + DESI + PPS dataset. On the other hand, the $\Delta$IC values derived from the CMB, DESI, PPS and CMB + DESI datasets show good alignment between $\Lambda$CDM and the EPN models with $M=-1/3$ and with adjustable $M$, although the former is mildly tensioned when considering combined CMB + DESI + PPS data.\footnote{The favorable alignment between $\Lambda$CDM and the special EPN model with $M=-1/3$ inferred from DESI data is in agreement with the analysis of \cite{Sudharani:2024qnn}.} The CMB + DESI dataset appears to favor both of these models over $\Lambda$CDM. Moreover, the tension probes suggest that CMB and DESI data are significantly more compatible in these EPN set-ups than in $\Lambda$CDM. The model with $M$ as fitting parameter also shows an improved credibility from the CMB + DESI + PPS dataset resulting in a slight alleviation of the tension between CMB + DESI and PPS data.

The improved compatibility between the CMB and DESI datasets can be seen in the CMB power spectra, as shown in the reconstructed high-$\ell$ TT power spectrum shown in Fig.\ \ref{fig: High-l Comspec TTTEEE power_spectrum}: there are evident discrepancies between the high-$\ell$ TT power spectra in $\Lambda$CDM fitted against CMB data and against CMB + DESI and CMB + DESI + PPS data, which is visibly alleviated in both of these realizations of EPN. For completeness, we also display the low-$\ell$ TT power spectrum in Fig.~\ref{fig: Low-l Commander power_spectrum}. It is clear that, as long as $c_m \gg 1$, EPN matches closely with $\Lambda$CDM, with a very slight suppression for the lowest $\ell$ modes. We remark that such a suppression cannot explain the quadruple anomaly~\cite{Planck:2019evm}. 

Fig.~\ref{fig: DL-z} shows the reconstructed distance modulus obtained from various dataset combinations. The manifest difference between the resulting fits provides an explanation for the tension between CMB + DESI and PPS data. The results for the $\chi^2_{\rm SN}$ distribution also shown again the mild preference of the special EPN model with $M$ as adjustable parameter. Interestingly, while models with fixed $M$ perform worse than $\Lambda$CDM in regards to this tension, the more minimal and natural set-up with $M=-1/3$ shows a strong preference over other theoretically motivated choices.


\section{Conclusions}
\label{sec: conclusion}

The aim of this work is to evaluate the viability of the vector-tensor EPN theory as a consistent and well-motivated framework of dynamical dark energy and identify whether it may be competitive against the standard $\Lambda$CDM model (accounting for an increase in parameter space). Our first observation was that EPN accommodates an interesting one-parameter extension of background cosmology, encompassing the so-called special (minimally coupled) EPN model, particular cases of the general EPN model, as well as other theories such as GP. This extension corresponds to an effective dark energy fluid with density $\rho_{\rm EPN}\propto H^{2M}$, with $M$ a parameter that may be freely adjusted in phenomenological studies such as the present one. On the other hand, it is an appealing feature that such modifications may emerge in EPN theory without the inclusion of operators that are unnatural from an EFT perspective, although it must be recalled that one still needs some degree of fine tuning in order to obtain simple decoupled equations for the background functions. We think our results motivate the analysis of more general parameterizations in future studies, including ones arising from models without any fine-tuning. At the same time, the degeneracy of different theories at the background level motivates the inclusion of perturbations when confronting the predictions of EPN with data. Here we have taken the first step in this direction, by analyzing the fit of the matter power spectrum and structure growth rate, within a particular version of the EPN special model.

The main target of our analysis was to assess whether the EPN theory (or akin theories) can have the potential to address the notorious tensions currently present in cosmological data. We have found that, in a particular realization, the special EPN model results in an easing of the Hubble tension from $5.81\sigma$ to $1.52\sigma$, but at the cost of an increase in the $\Om{m0}$ tension from $0.85\sigma$ to $5.19\sigma$, netting an increase in all tension probes utilized in this work. This is consistent with other thawing phantom models, in particular the CPL phenomenological model with a matching cosmographic equation of state. 

Our results also highlight the critical importance of perturbations when confronting the model with current data, improving on previous work that focused exclusively on the background dynamics. We find that `natural' values of the Proca mass, i.e.\ $m\sim \Lambda^2/\MP$, result in significant enhancements of CMB foreground effects. Agreement with data can however be recovered  at ever so slightly larger masses, resulting in parameter posteriors, growth history, and matter power spectrum which are similar to the toy set-up where dark energy perturbations are turned off. We emphasize that this regime is acceptable both phenomenologically and theoretically, as it amounts to a mild  tuning in only one model parameter. As long as stability criteria are met, this set-up thus provides a consistent description which allows one to isolate the observational effects that result from the modification of the Friedmann equation. This is very pertinent since, as we have emphasized repeatedly, the same class of modification is common to several other dark energy models with vector degrees of freedom. The cosmological fit of this background-only one-parameter special M model shows that it is favored over $\Lambda$CDM by $1.5\sigma$ when combining \textit{Planck} datasets with DESI BAO, and by $2.4\sigma$ when further adding the PantheonPlus+SH0ES dataset. We also note that our analysis shows no evidence of an upper limit on the allowed value of the mass.

Returning to the full set-up with perturbations, we find that the theoretically motivated very special EPN model also performs better than $\Lambda$CDM when comparing the derived $S_8$ value of the DES Y3 weak lensing dataset with the predicted matter perturbation power at $8\,\text{Mpc}/h$, thus easing the $S_8$ tension. However, when analyzing the complete matter power spectrum, we notice a strong enhancement at the BAO scale, in conflict with the full-shape analysis of the SDSS galaxy clustering dataset. This large-scale matter power enhancement, which cannot be captured by $\sigma_8$, 
may therefore be seen as an important metric for assessing the viability of EPN theories beyond the background level. The growth history of this particular model is also marginally different from that of $\Lambda$CDM. In combination, the large-scale structure dataset may serve as a consistency check for EPN theory and motivates follow-up research on more general EPN models fully including perturbations.



In conclusion, we have found that EPN theory restricted to a one-parameter subclass of models performs marginally better than $\Lambda$CDM, although at the price of a mild tuning of parameters which suppresses the effect of dark energy perturbations. Moreover, it is clear that this particular framework is unsuccessful at resolving all cosmological tensions, yet we find our results to be promising enough to warrant further studies that include a wider parameter space, which could investigate the possibility of additional mechanisms for altering the evolution of matter perturbations and potentially address the aforementioned tensions.

\acknowledgments

We would like to thank Lavinia Heisenberg for useful discussions. The work of HWC, SGS and XZ was partly supported by the NSFC (Grant No.\ 12250410250). The work of CdR is supported by STFC Consolidated Grant ST/X000575/1. CdR is also supported by a Simons Investigator award 690508. SGS also acknowledges support from a Provincial Grant (Grant No.\ 2023QN10X389).

\appendix

\section{Initial conditions, effective Newton constant, and anisotropic stress}
\label{sec: CAMB}

\subsection{Adiabatic initial conditions}

To perform the numerical integration of the perturbation equations in \texttt{CAMB} we adopt the standard choice of adiabatic initial conditions. The task is therefore to identify the adiabatic modes for the EPN system considered in this paper. To this end, and in what follows, it proves convenient to introduce the gauge-invariant quantity
\begin{equation}
\delta_{B,i} \equiv \frac{\delta \rho_i}{\rho_i} + \frac{3H}{\MP^2} (1+w_i) v_i \,,
\end{equation}
for each matter species $i$, as well as the rescaled (dimensionless) variables
\begin{equation}
\tilde \psi \equiv \frac{H}{\Lambda\phi} \psi  \,,\qquad
\tilde{\mathcal Y} \equiv \frac{1}{\MP^2\Lambda^2\Omega_{\rm EPN}} \hat{\mathcal Y} \,.
\end{equation}
One advantage is that the master equations for these variables, when expressed in terms of conformal time, depend only on the ratio between the comoving wavenumber and the horizon scale, $k/(a H)$, and do not contain poles at the radiation, matter, or de Sitter fixed points, allowing us to unambiguously and properly define the sub-horizon and super-horizon modes in each era.

To a good approximation, we may simplify the system by assuming a matter sector composed of cold dark matter `${\rm c}$' and radiation `${\rm r}$'. In this set-up one may easily identify the only mode (with degeneracy two) that is non-decaying before the dark energy-dominated era ($\Omega_\EPN \to 0$). We expect these results to be applicable to the four-fluid system (which also includes baryons and neutrinos) implemented in \texttt{CAMB}.

We write the perturbation equations as a first order matrix system in the basis $\{\delta_{B,\rm c},\delta_{F,\rm c}$, $\delta_{B,\rm r},\delta_{F,\rm r},\tilde{\mathcal Y},\tilde \psi\}$, where $\delta_{F,i}\equiv \delta\rho_i/\rho_i$ is the density contrast (the subscript stands for `flat', as this quantity is gauge-dependent and we remind the reader that we are working in the spatially flat gauge). We then find the following set of eigen-modes:
\begin{align}
&\begin{array}{ccccccccc}
E_{{\rm c},+}         &= \Big(\!&  \Delta                                 \,,\!&  3 + \Delta      \,,\!&  0           \,,\!&  4       \,,\!&  0       \,,\!&  1                       \!&\Big)\,,  \\[6pt]
E_{{\rm c},-}         &= \Big(\!&  \frac{4 \Om{r}}{\Om{c}} + 3 + \Delta   \,,\!&  3 + \Delta      \,,\!&  0           \,,\!&  4       \,,\!&  0       \,,\!&  1                       \!&\Big)\,,  \\[6pt]
E_{{\rm r},\pm}       &= \Big(\!&  0                                      \,,\!&  3               \,,\!&  \frac{1 + \Om{r}}{2\Om{r}} (1 \pm i \omega_J ) + \frac{\Om{c}}{\Om{r}} \Delta   \,,\!&  4 + \frac{\Om{c}}{\Om{r}}\Delta     \,,\!&  0       \,,\!&  1       \!&\Big)\,,  \\[6pt]
E_{\EPN, \pm} \!&= \Big(\!&  0                                      \,,\!&  0               \,,\!&  0           \,,\!&  0       \,,\!&  1       \,,\!&  \tilde \psi_{\EPN, \pm} \!&\Big)\,,
\end{array}\label{eq: eigen modes}
\end{align}
with respective decay rates (i.e.\ minus the eigenvalues)
\begin{align}
\frac{\lambda_n}{H} = \left( 0\;,\; \frac{3 + \Om{r}}{2}\;,\; \frac{1 + \Om{r}}{4} \left(1 \pm i \omega_J\right)\;,\; \frac{1}{4} \bigg{(} 7 + 3 \Om{r} - 2M ( 3 + \Om{r} )  \right. \nonumber\\
\left. \mp \sqrt{\left( 5 + \Om{r} \right)^2 + 4 M (1 - \Om{r}) (5 + \Om{r} - M ( 3 + \Om{r}) ) \Delta} \bigg{)} \right)  \,,
\end{align}
where $\Delta \equiv \tfrac{2}{3\Omega_{\rm c}}\tfrac{k^2}{a^2H^2}$ and $\omega_J^2 \equiv k^2/k_J^2 - 1$ measures how much a mode $k$ lies below the Jeans scale $k_J^2 \equiv \tfrac{3}{16} a^2 H^2 \left( 1 + \Om{r} \right)^2$. We also introduced
\begin{equation}
\begin{aligned}
\tilde \psi_{\EPN, \pm} &\equiv -\frac{1}{8M ( 3 + \Om{r} + ( 1 - \Om{r} ) \Delta )} \bigg[ 5 + \Om{r} + 2M ( 1 - \Om{r} ) \Delta  \\
&\quad \pm\sqrt{ \left( 5 + \Om{r} \right)^2 + 4 M ( 1 - \Om{r} ) ( 5 + \Om{r} - M (3 + \Om{r}) ) \Delta} \bigg]  \,,  
\end{aligned}
\end{equation}

On super-horizon scales ($k \ll a H$) we have $\omega_J \simeq i$ and $\Delta \simeq 0$. We then see that the only frozen modes are $E_{{\rm c},+}$ and $E_{{\rm r},+}$. In the strict limit $k=0$, these modes are degenerate and equal to $\left(0\,,\,3\,,\,0\,,\,4\,,\,0\,,\,1\right)$, which corresponds to the standard adiabatic mode, which sets equal amplitudes for the perturbations $\tfrac{1}{3}\delta\rho_{\rm c}/\rho_{\rm c}$, $\tfrac{1}{4}\delta\rho_{\rm r}/\rho_{\rm r}$ (i.e.\ $\tfrac{1}{3(1+w_i)}\delta\rho_i/\rho_i$) and $\tilde{\psi}$, with zero $\delta_{B,i}$ and $\tilde{\mathcal Y}$. Of course, away from this strict limit, there exists an orthogonal combination, $E_{{\rm c},+}-E_{{\rm r},+}$, which however has $\delta_{B,{\rm c}}=\delta_{F,{\rm c}}$, i.e.\ $v_{\rm c}=0$. Such mode thus cannot be sourced by the curvature perturbation, which sets $v_i\neq0$ for all species on super-horizon scales \cite{Weinberg:2003sw}.

Having identified the adiabatic mode, we may then match its super-horizon expression to the corresponding one on sub-horizon scales ($k \gg a H$).\footnote{At this stage we notice an obvious issue with our choice of basis, namely that $\Delta\to\infty$ in the sub-horizon limit, leading to an artificial hierarchy between $E_{\EPN,\pm}$ and the rest of the eigen-modes. We can easily remedy this by rescaling $\tilde \psi$ and $\tilde{\mathcal{Y}}$ by $\Delta$. This redefinition introduces a mixing between $E_{{\rm c},+}$ and $E_{\EPN,\pm}$ (but not with the other modes), however this is irrelevant for our purposes since the latter is always decaying (provided the stability criteria are met).} We focus on the ratio $\delta_{B,{\rm c}}/\tilde \psi$ of the mode $E_{{\rm c},+}$, the advantage being that this quantity is independent of the coupling between photons and baryons, and thus robust in our approximation where we neglect the latter. Thus we conclude that the choice of adiabatic initial conditions is given by $\tilde \psi = \Delta^{-1} \delta_{B,{\rm c}}$ and $\tilde{\mathcal{Y}} = 0$ (with $\delta_{B,{\rm c}}$ evaluated in synchronous gauge for usage in \texttt{CAMB}).

\subsection{Effective Newton constant, anisotropic stress and integrated Sachs-Wolfe effect}

\begin{figure}
\centering
\subfigure[Effective Newton constant \label{fig: cmb Geff}]{\includegraphics[width=0.49\linewidth]{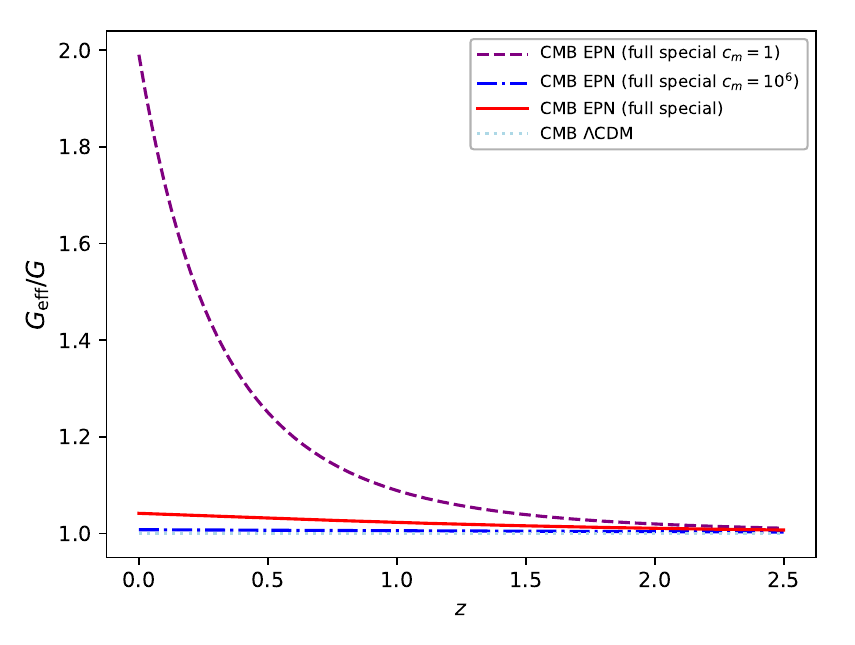}}~
\subfigure[ISW-galaxy cross-correlation \label{fig: cmb F}]{\includegraphics[width=0.49\linewidth]{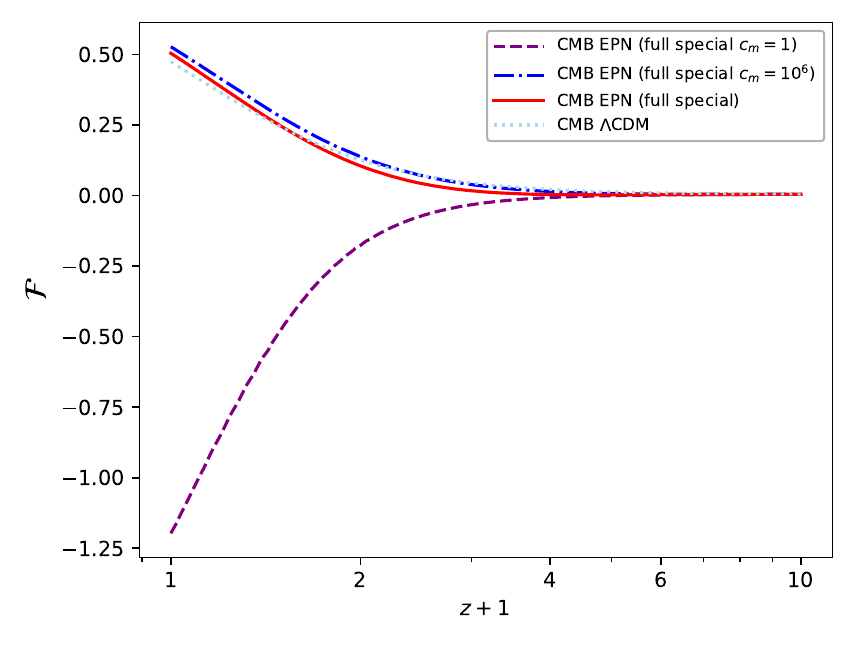}}
\caption{Evolution of the effective Newton constant $G_{\rm eff}$ (\emph{left panel}) and the ISW-galaxy cross-correlation measure $\mathcal F$ (\emph{right}), in the best-fit $\Lambda$CDM model and the EPN full special model given the best-fit parameters for the CMB dataset, utilizing different values of the Proca mass parameter $c_m$, including the best-fit value $c_m=10^{4.48}$ ({\color{red}red} curve). The strong enhancement of $G_{\rm eff}$ and the negativity of $\mathcal F$ for $c_m = \mathcal{O}(1)$ values are manifest.}
\end{figure}

The anisotropic stress $\Pi$ and effective Newton constant $G_{\rm eff}$ are defined via
\begin{equation}
\begin{aligned}
\Pi &\equiv \frac{3a^2}{\MP^2}\left(\frac{k^i k^j}{k^2} - \frac{1}{3}\delta^{ij} \right)T_{ij} = 2k^2 ( \Psi_B + \Phi_B ) \,,\\
-2k^2\Psi_B &\equiv\frac{G_{\rm eff}}{G}\frac{a^2}{\MP^2} \sum_i \rho_i \delta_{B,i} \,,
\end{aligned}
\end{equation}
where $\Psi_B$ and $\Phi_B$ are the Bardeen potentials. In our conventions, these are given in terms of the spatially flat gauge metric perturbations by
\begin{equation}
\Psi_B = \frac{1}{\MP} \alpha + \frac{1}{\MP^2} \dot\chi \,,\qquad \Phi_B = \frac{H}{\MP^2}\chi \,.
\end{equation}
Through some straightforward manipulations of the perturbation equations of Sec.\ \ref{subsec: Evolution of the perturbation in the minimally coupled EPN model}, we eventually obtain
\begin{align}
\Pi &= \frac{9a^2 H^2}{\MP^2} \sum_i \Omega_i\left[ 3H ( c_i^2 - w_i ) ( 1 + w_i ) + \dot w_i \right] v_i  \,,\\
\frac{G_{\rm eff}}{G} &= 1 + \frac{\MP^2}{a^2 \sum_i \rho_i \delta_{B,i}} \left( \frac{k^2}{2} \tilde{\mathcal{Y}} - \Pi \right)  \,.
\end{align}
Given $c_i^2 = w_i={\rm const.}$ under the perfect fluid approximation, we conclude that $\Pi = 0$ and $G_{\rm eff}/G = 1 + k^2 \tilde{\mathcal{Y}} / ( 6 a^2 H^2 \sum_i \Omega_i \delta_{B,i})$. The vanishing of the anisotropic stress agrees with \cite{DeFelice:2016uil,deFelice:2017paw} which derived the same result in the context of GP theory. Fig.~\ref{fig: cmb Geff} displays how $G_{\rm eff}/G$ starts deviating from unity during the matter-dark energy transition, consistent with the adiabatic assumption on the initial conditions, $\tilde{\mathcal{Y}} \to 0$, and the fact that $\tilde{\mathcal{Y}}$ remains frozen prior to the transition.

The fact that $G_{\rm eff}/G\neq 1$ inevitably affects the integrated Sachs-Wolfe (ISW) effect, i.e.\ the spectral distortion due to CMB photons gaining or losing energy as they climb in and out of the evolving gravitational potential. Since the ISW effect is sensitive to the time dependence of the potential, one can utilize the cross-correlation between the ISW part of the CMB spectrum and the CDM density perturbation to isolate the quantity $\mathcal F \equiv - \frac{d\ln\Psi_B}{d\ln a}$~\cite{Nakamura:2018oyy}. This observable has been measured to be positive~\cite{Stolzner:2017ged}, indicating a positive cross-correlation, as predicted by $\Lambda$CDM. Dynamical dark energy models may however predict the opposite behavior (see e.g.~\cite{Kable:2021yws}), although vector-tensor theories generally alleviate this issue~\cite{Nakamura:2018oyy}. This is confirmed in EPN: although values of $c_m = \mathcal O (1)$ indeed predict a negative cross-correlation, mildly large values (including the special model best-fit choice) restore consistency, with $\mathcal F$ matching the $\Lambda$CDM behavior with reasonable approximation. We remark however on the slight enhancement of $\mathcal F$ even in the limit of large $c_m$, which must therefore be ascribed to the difference in the background evolution between $\Lambda$CDM and EPN.

\section{Information Criteria}
\label{sec: Information Criteria}

The following is a list of the definitions of the information criteria probes utilized in this work:
\begin{align}
\ln B (D|M) &\equiv \ln \int P(D|\theta_M) P(\theta_M) d\theta_M = -\ln V_M - \ln \left< \left( P(D|\theta_M) \right)^{-1} \right>_{M|D}  \label{eq: -lnB}  \,,\\
{\rm DIC} (D|M) &\equiv \ln P(D|\theta_{M,\,map}) - 2F(D|M)  \,,\\
{\rm WAIC} (D|M) &\equiv - F(D|M) + {\rm BMD} (D|M) / 2  \,,\\
F(D|M) &\equiv \left< \ln P(D|\theta_M) \right>_{M|D} \equiv \ln B (D|M) + {\rm KL} (D|M)  \,,\\
{\rm BMD} (D|M) &\equiv 2 \left( \left< \left( \ln P(D|\theta_M) \right)^2 \right>_{M|D} - \left< \ln P(D|\theta_M) \right>_{M|D}^2 \right)  \,,\\
-\ln R (D_1, D_2 | M) &\equiv - \ln B (D_1 D_2 | M) + \ln B (D_1 | M) + \ln B (D_2 | M)  \,,\\
{\rm GoF} (D_1, D_2 | M) &\equiv - \ln P(D_1 D_2 |\theta_{M,\,map}) + \ln P(D_1 |\theta_{M,\,map}) + \ln P(D_2 |\theta_{M,\,map})  \,,\\
S (D_1, D_2 | M) &\equiv - F(D_1 D_2 | M) + F (D_1 | M) + F (D_2 | M)  \,.
\end{align}
Here $D$ and $M$ denote respectively the dataset and the model, and $\theta_M$ stands for the model parameters; $P(D|\theta_M)$, $P(\theta_M)$, $P(\theta_M|D) \equiv P(D|\theta_M) P(\theta_M) / B(D|M)$ are the likelihood, prior probability and posterior probability, respectively; $\left< \ldots \right>_{M|D} \equiv \int (\ldots) P(\theta_M|D) d\theta_M$ is the MCMC mean, $V_M \equiv \int P(\theta_M) d\theta_M$ is the prior volume, $F$ is the deviance, and ${\rm KL} (D|M) \equiv \left< \ln P(\theta_M|D) - \ln P(\theta_M) \right>_{M|D}$ is the Kullback-Leibler divergence; BMD stands for the Bayesian model dimension, and $map$ is the abbreviation of maximum-a-posteriori, i.e.\ the Bayesian estimate of the model parameters. Our choice of normalization is such that $\ln B$, ICs, $\ln R$, GoF and suspiciousness $S$ all have the same comparative measure as the Bayesian ratio test, i.e.\ the Jeffreys scale measures; cf.\ Table~\ref{table: Jeffrey’s scale}.

\begin{table}[]
\centering
\begin{tabular}{cc}
\toprule
\midrule
$\ln B (M_2) - \ln B (M_1)$, $\ln R$, etc.    & Interpretation    \\
\midrule
$>5$        & Strongly disfavored / tensioned   \\
$2.5\sim5$  & Moderately disfavored / tensioned \\
$1\sim2.5$  & Weakly disfavored / tensioned     \\
$-1\sim1$   & Inconclusive                       \\
$-2.5\sim-1$& Weakly favored / aligned          \\
$-5\sim-2.5$& Moderately favored / aligned      \\
$<-5$       & Strongly favored / aligned        \\
\midrule
\bottomrule
\end{tabular}
\caption{\label{table: Jeffrey’s scale}
Jeffreys' scale for evaluating the evidence of model $M_1$ over $M_2$ or the tension between datasets.
}
\end{table}

\newpage

\section{Full constraint plot of cosmological parameters} \label{sec:appendix full constraint plot}

\begin{figure}[h]
\centering
\includegraphics[width=1\linewidth]{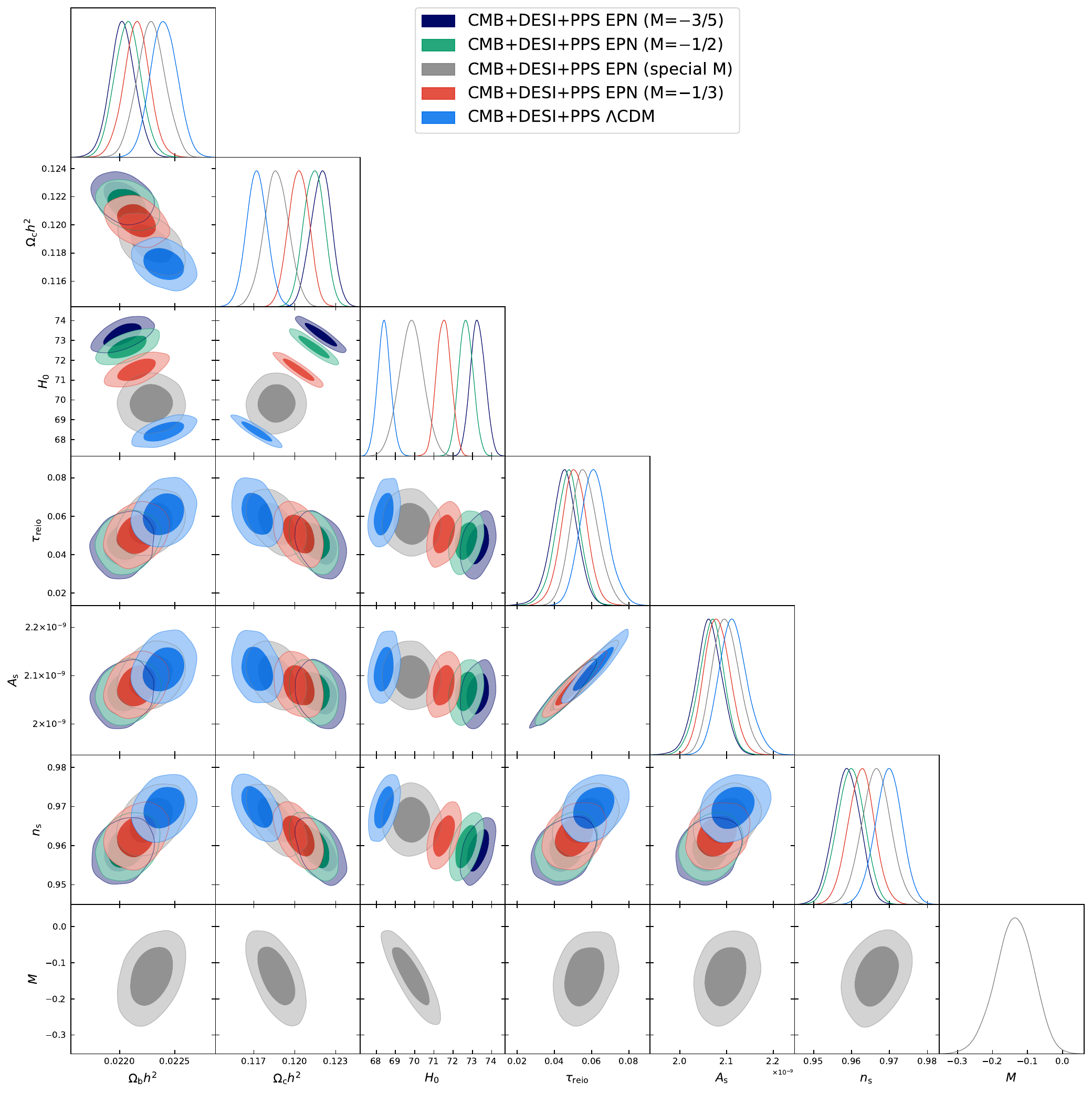}
\caption{Marginalized posterior constraints on the six base cosmological parameters and special EPN model parameter $M$, using the CMB+DESI+PPS dataset, in $\Lambda$CDM, three EPN models with fixed $M$, and EPN special M model with unfixed $M$. Contours indicate 68\% and 95\% confidence intervals. The resulting constraint on $M$ from this dataset is $ M=-0.136^{+0.057}_{-0.051}$ (68\% C.L.).}
\label{fig: contour_CMB+DESI+PPS 6 parameters}
\end{figure}

\newpage

\bibliographystyle{JHEP}

\bibliography{bibliography.bib}

\end{document}